\documentclass[showpacs,twocolumn]{revtex4}

\usepackage{graphicx}
\usepackage{epsfig}             

\begin{document}

\title{Characterization of EIT-based continuous variable quantum memories}

\author{G.~H\'etet, A.~Peng, M.~T.~Johnsson, J.~J.~Hope, P.~K.~Lam}
\email[Email: ]{ping.lam@anu.edu.au}
\affiliation{Australian Centre for Quantum-Atom Optics, Department
of Physics, Australian National University, ACT 0200, Australia}

\date{\today}

\begin{abstract}
We present a quantum multi-modal treatment describing Electromagnetically Induced Transparency (EIT) as a mechanism for storing continuous variable quantum information in light fields.  Taking into account the atomic noise and decoherences of realistic experiments, we model numerically the propagation, storage, and readout of signals contained in the sideband amplitude and phase quadratures of a light pulse. 
An analytical treatment of the effects predicted by this more sophisticated model is then presented.
Finally, we use quantum information benchmarks to examine the properties of the EIT-based memory and show the parameters needed to operate beyond the quantum limit.
\end{abstract}


\maketitle

One of the steps towards the realization of quantum computation is a device that allows the 
coherent storage of information. The Heisenberg Uncertainty Principle (HUP) sets a limit on 
the quality of stored information that depends on direct measurement and subsequent 
reconstruction. Much experimental and theoretical research is 
directed towards quantum memories for light to circumvent this classical benchmark.  
To realize such memories, methods that provide a coherent interface between 
large atomic ensembles and light fields have been proposed.

A scheme using the off-resonant interaction of a light field with a large ensemble of three level atoms was presented in \cite{Koz}. 
Off-resonant Faraday rotation was also used as a mechanism for mapping quantum states of light onto atoms \cite{Juls04}. 
The storage of a light field was shown to be possible by controlling the spatial distribution of atomic shifts in optically thick ensembles of three level \cite{Moiseev,Sangouard} and two level atoms \cite{Hetet}.  Probably the most actively studied technique to achieve a quantum memory for light utilizes Electromagnetically Induced Transparency (EIT) \cite{Liu,fleischhauer}. 

Experiments using EIT in atoms carried out in a sodium magneto-optical trap (MOT) 
\cite{Liu} and in hot rubidium vapor cells 
\cite{phillips} have demonstrated the storage of a light pulse for a few milliseconds. In solid state systems a storage of more than 1s has been achieved
using photon echo techniques \cite{longdell}. 
The quantum nature of single photon Fock states was shown to be preserved when stored and released from a MOT \cite{chaneliere, laurat, Eisaman}, and theoretical studies have proposed methods to enhance the storage efficiency in these experiments \cite{Gor, dantan1,Nunn}.

Although controlled storage of the amplitude and phase quadratures of a probe field at the quantum limit using 
EIT has not been achieved, experimental efforts in this direction have shown the delay of the two quadratures of a continuous wave beam
\cite{hsu} and the transmission and delay of vacuum squeezing
\cite{akamatsu,akamatsu2}.  Improvements in the efficacy of this system can be made with a better understanding of the sources of excess noise and loss.  The transfer of the sideband statistics from optical fields to atoms also requires further investigation.  In this work we develop a model describing the storage of the signals contained in the sideband amplitude and phase quadratures of a light pulse in the presence of decoherences and associated atomic noise, and use quantum information benchmarks to show the quantum nature of the transfer.

In the first part we present theoretical models that describe the multi-mode propagation of an amplitude and phase modulated pulse and the storage of its information onto atomic states in EIT-based memories. A numerical phase space treatment of light storage treats several sources of inefficiency present in current experiments. Linearized Maxwell-Bloch equations are then solved analytically in the weak probe approximation to explain the behaviour of the atomic noise and to give an expression for the time-bandwidth product of this system in the presence of decoherence and finite atomic density.

Next, we develop criteria that quantify parameters for which EIT based memories are able to store information in the quantum regime.
Several criteria have been developed in the past to distinguish classical and quantum
distributions of states in other quantum information protocols, such as teleportation or quantum cryptography.
Signal-transfer coefficients $T$, and conditional variances $V_{\rm cv}$, have been used as a state independent measure 
to analyze the effectiveness of teleportation experiments in the presence of non-unity gain \cite{ralph, ralph2, bowen}.
We propose to implement the {\it TV diagram} to define benchmarks for the storage of continuous variable information
and identify the parameters required to enable a transfer of information that outperforms any classical strategy.

\section{Model}
   
Previous theoretical work has characterized the efficiency of EIT as a delay line for continuous variable quantum states \cite{peng, dantan1}. 
Considering a three level atom, under conditions where there is a pure dephasing rate between the ground states, information 
can be slowed down within a narrow frequency window, and no additional noise is introduced beyond that which is necessary to preserve the canonical commutation relation of the field \cite{peng}.
The width of the transparency window depends on the coupling beam power and the atomic density.
Controlling the coupling beam in time allows storage of the information within the atomic sample.

This storage process can be understood as follows.
The coupling beam prepares the atoms initially in state $|1\rangle$ through optical pumping. When a weak probe propagates in the medium under EIT conditions, 
coherences are created between the two ground states of the atoms. 
These coherences arise from a quantum interference between the two fields and acquire the sideband information
 of the probe pulse during its compression inside the medium. After the compression, most of the probe field energy has been transferred to 
 the coupling beam and left the cell at the speed of light. At this point in time the atoms possess the frequency information of the 
 probe within the transparency window, distributed in momentum space. 
When the coupling beam is turned off, the remaining energy in the probe field leaves the medium without affecting the information stored.
The information will be saved provided the readout is performed before the decoherence processes 
have affected the atomic state. When the coupling beam is switched back on, the probe beam is regenerated with the supply of photons from the coupling 
beam and leaves the medium while reading the spin state of the atoms. The main constraints are that the signal has to be encoded at frequencies
 within the transparency window and that the compressed pulse has to fit within the sample size. When these criteria are satisfied, the efficiency of this process is close to unity.
 
Experimental investigation of this effect requires optical sources at or below the shot noise limit, which is only possible at some modulation frequency around a carrier.  Modeling this spatio-temporal quantum information accurately therefore requires a model which contains the quantum state of a large number of modes of the light, which we provide in this paper.  We solve this problem numerically and then analytically to calculate the degradation of the signal and added noise during the storage process in the presence of decoherence mechanisms. Specifically, we consider dephasing affecting the ground state coherence and also allowing an exchange of population between the two ground states.

\begin{figure}[!ht]
\begin{center}
\includegraphics[width=5cm]{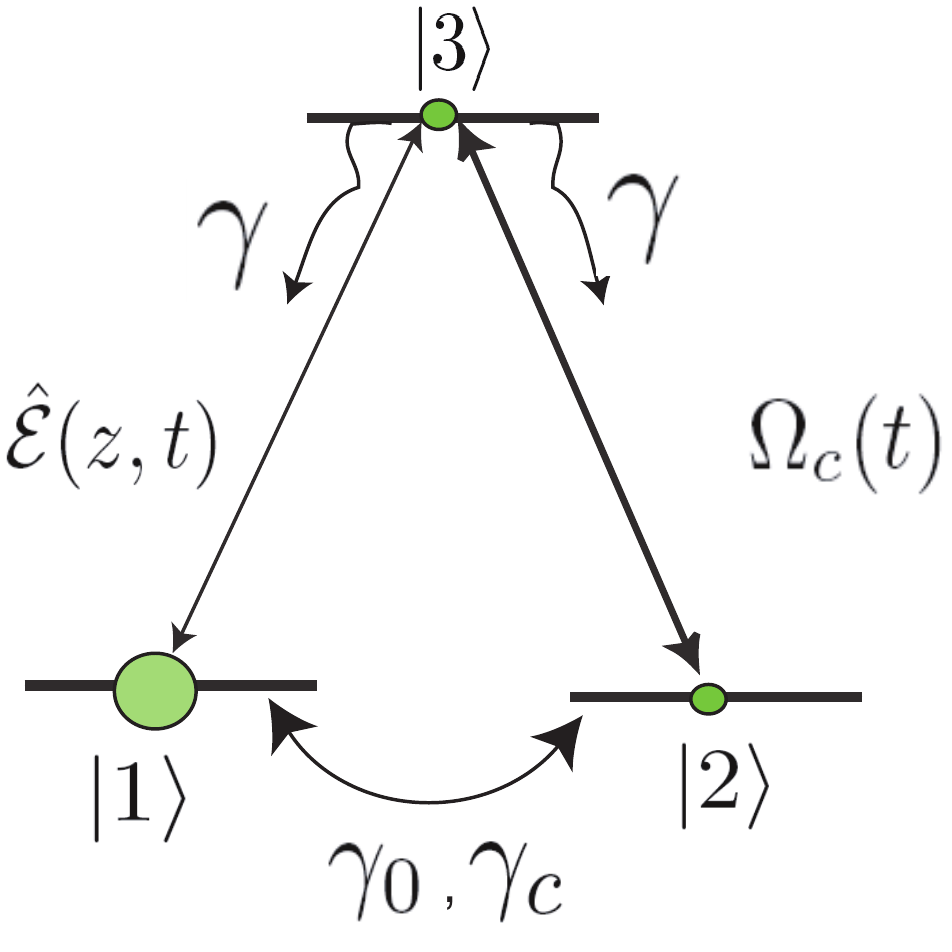}
\caption{EIT level structure. $\hat{\mathcal{E}}(z,t)$ is the envelope operator of the probe field, and $\Omega_{c}(t)$ the coupling beam Rabi frequency. 
Almost all the atoms are pumped into state $|1\rangle$ initially. $\gamma$ is the spontaneous emission rate from the upper state and $\gamma_{0}$, $\gamma_{c}$ are mean decoherence rates between the two ground states for pure dephasing and population exchange respectively. These two quantities are usually referred to as $1/T_{2}$ and $1/T_{1}$ in the field of magnetic resonance.}
\label{levels}
\end{center}
\end{figure}

We approximate the atomic structure by the three level atomic Lambda system shown in Fig.~\ref{levels},
where the two atomic ground states are degenerate and the transitions are addressed experimentally with orthogonal circular polarizations.
The switching of the coupling beam can be done adiabatically or abruptly if the pulse is totally compressed within the medium \cite{matsko2, zibrov, Liu}.
Here we only consider the case where the coupling beam is switched abruptly, although other theoretical works show that the way the coupling beam is 
shaped in time enhances the efficiency \cite{Gor, Nunn}.

In the following calculation, we consider the simultaneous storage of both quadratures of the probe when amplitude and phase modulations are encoded within the EIT bandwidth. The preparation of the state can be achieved experimentally by passing a light pulse through amplitude and phase modulators sequentially.
Provided the modulation frequency is larger than the Fourier width $\Delta \omega$ of the pulse, 
classical information is encoded onto its sideband $\omega$ at the shot noise limit.

We are interested in the envelope of the probe field, so the problem will be solved using the rotating wave approximation.
The envelope operator is denoted by $\hat{\mathcal{E}}(z,t)$, and its commutator is $[\hat{\mathcal{E}}(z,t),\hat{\mathcal{E}}(z',t')]=\frac{L}{c}\delta (t-z/c-(t'-z'/c))$,
where $L$ is the quantization length, taken to be the length of the cell and $c$ is the speed of light.
Let $\hat{X}^{\pm}_{\rm in}(\omega)$ be the quadrature operators of the input probe field at the sideband frequency $\omega$.
For the amplitude and phase quadratures we have $\hat{X}^{+}_{in}(\omega)=\hat{\mathcal{E}}_{\rm in}(\omega)+\hat{\mathcal{E}}^{\dagger}_{\rm in}(-\omega)$ and $\hat{X}^{-}_{in}(\omega)=-i(\hat{\mathcal{E}}_{\rm in}(\omega)-\hat{\mathcal{E}}^{\dagger}_{\rm in}(-\omega))$, respectively.

We can write $\hat{X}^{\pm}_{\rm in}(\omega)=2\alpha^{\pm}_{\rm in}(\omega)+\delta \hat{X}^{\pm}_{\rm in}(\omega)$,
where $\alpha_{\rm in}^{\pm}(\omega)$ is the coherent amplitude encoded onto the probe via optical modulation, and $\delta \hat{X}^{\pm}_{in}(\omega)$
its quantum fluctuations. 
The power spectral density $S^{\pm}(\omega)$ of a signal is the Fourier transform of the autocorrelation function and obeys the relation \cite{peng,dantan2}

\begin{eqnarray}
S^{\pm}(\omega)\delta(\omega+\omega')=\frac{c}{L}\langle\hat{X}^{\pm}(\omega)\hat{X}^{\pm}(\omega')\rangle\label{powspect}.
\end{eqnarray}
When normalized to the detection bandwidth, chosen to be much smaller than the applied modulation frequency,  the measured power spectrum is 
\begin{eqnarray}
S^{\pm}(\omega)=\frac{c}{L}\langle |\hat{X}^{\pm}(\omega)|^2 \rangle.
\end{eqnarray}

We also need to define the noise floor of the signal by introducing the measured power spectrum without signal 

\begin{eqnarray}
V^{\pm}(\omega)=\frac{c}{L}\langle |\hat{\delta X}^{\pm}(\omega)|^2 \rangle.
\end{eqnarray}

For the input probe state we then have $S^{\pm}_{\rm in}(\omega)=4\frac{c}{L}(\alpha_{\rm in}^{\pm}(\omega))^{2}+V^{\pm}_{\rm in}(\omega)$.
The signal is defined as $4\frac{c}{L}(\alpha_{\rm in}^{\pm}(\omega))^{2}$ and the noise as $V^{\pm}_{\rm in}(\omega)$, which is unity for a shot noise limited laser beam.

If this state is inefficiently stored with some frequency and quadrature dependant linear loss $\eta^{\pm}(\omega)$ and if some excess noise with amplitude $V^{\pm}_{\rm noise}(\omega)$ is generated by the memory, we have 
\begin{equation}
S^{\pm}_{\rm out}=\eta^{\pm}(\omega) S^{\pm}_{\rm in}+1-\eta^{\pm}(\omega)+V^{\pm}_{\rm noise}(\omega).
\end{equation}

The term $1-\eta^{\pm}(\omega)$ corresponds to uncorrelated vacuum noise, common to any system in the presence of linear loss.
This noise term preserves the purity of a quantum state and is necessary to preserve the commutation relations of the output state.
On the other hand, a device generating some excess noise $V^{\pm}_{\rm noise}(\omega)$, transforms
an initial coherent state, $V^{\pm}_{\rm in}(\omega)=1$ into a mixed state where $V^{\pm}_{\rm out}(\omega)=1+V^{\pm}_{\rm noise}(\omega)>1$.
In the following section we will calculate $\eta^{\pm}(\omega)$ and $V^{\pm}_{\rm noise}(\omega)$ using phase space simulations in the positive P 
representation.

\subsection{Stochastic Simulations}
    
We treat the probe beam as a general quantized field with longitudinal spatial dependence $z$, and the coupling beam $\Omega_c(t)$ as a classical field.
The atoms are all prepared in state $|1\rangle$ before the probe enters the cell via optical pumping induced by $\Omega_{c}$. 
In this study we assume the coupling beam Rabi frequency to be $10^4$ times larger than that of the probe,
ensuring that no atoms will move into state $|2\rangle$ due to optical pumping induced by the probe and the coupling beam will not be depleted throughout the storage process. 
Its dependence on $z$ will therefore be ignored in this treatment.
The validity of this approximation in the presence of decoherence is discussed in the next section.
When both beams are of comparable strength and are both treated as quantum fields, a strong correlation also takes place between them 
\cite{martinelli, sautenkov}. 
This correlation will not affect the quantum statistics of the probe in our study so the coupling beam can be treated as a classical field. 

The master equation of this system is 

\begin{equation}\label{master}
\frac{\partial}{\partial t} \hat{\rho}= \frac{1}{i\hbar} [\hat{\mathcal{H}}_{\rm int},\hat{\rho}]+ \mathcal{L}_{31}[\hat{\rho}]+\mathcal{L}_{32}[\hat{\rho}]+
\mathcal{L}^{deph}_{[1,2]}[\hat{\rho}]+\mathcal{L}^{coll}_{[1,2]}[\hat{\rho}]
\end{equation}
where $ \hat{\rho}$ is the reduced density matrix of the field and atomic variables and $\hat{\mathcal{H}}_{\rm int}$ is the interaction Hamiltonian.
We define locally averaged atomic dipole operators $\hat{\sigma}_{ij}(z,t)$ for the $|i\rangle-|j\rangle$ transition given by \cite{fleischhauer,peng}
 
\begin{equation}
\hat{\sigma}_{ij}(z,t)=\frac{1}{n\mathcal{A} \delta z} \sum_{z_{k} \in \delta z} \hat{\sigma}^{k}_{ij}(z,t)
\end{equation}
where $\mathcal{A}$ is the cross-sectional area of the beam, $n$ the atomic density and $\delta z$ an infinitesimal slice of the medium containing $N$ atoms.
In the rotating wave approximation, the interaction Hamiltonian of the Lambda system is then
\begin{equation}
\hat{\mathcal{H} }_{\rm int} = - \int \frac{N \hbar}{L} [ g
\hat{\sigma}_{31}(z,t) \hat{\mathcal{E}}(z,t) + \Omega_c(t)
\hat{\sigma}_{32}(z,t) + H.c. ] dz \label{hamiltonian}
\end{equation}
where ${\rm g}$ is the coupling strength on the probe transition.
 The $\mathcal{L}_{i3}$ are Liouvillians modeling the decays due to spontaneous emission from the upper state $|3\rangle$, and are defined by
\begin{equation}
\mathcal{L}_{i3}[\hat{\rho}]=\gamma \sum_{z_{k} \in
\delta z}( \hat{\sigma}^{k} _{i3}\hat{\rho}\hat{\sigma}^{k} _{3i}-\frac{1}{2} \hat{\sigma}^{k} _{i3} \hat{\sigma}^{k} _{3i} \hat{\rho}-\frac{1}{2}\hat{\rho}\hat{\sigma}^{k} _{i3}\hat{\sigma}^{k} _{3i} )
\end{equation}
where for simplicity we assume the decay rates $\gamma$ from the upper state to be the same for both transitions.

$\mathcal{L}^{deph}_{[1,2]}$ accounts for an off-diagonal dephasing rate $\gamma_{0}$ affecting the ground state coherence and arises from elastic collisions or atoms moving in and out of the interaction region defined by the probe beam quantized mode. Its expression is
\begin{eqnarray}
\mathcal{L}^{deph}_{[1,2]}[\hat{\rho}]=\gamma_{0} \sum_{z_{k} \in \delta z} ( \hat{\sigma}^{k} _{11}\hat{\rho}\hat{\sigma}^{k} _{11}-\frac{1}{2} \hat{\sigma}^{k} _{11} \hat{\sigma}^{k} _{11} \hat{\rho}-\frac{1}{2}\hat{\rho}\hat{\sigma}^{k} _{11}\hat{\sigma}^{k} _{11}) \nonumber\\
+ \gamma_{0} \sum_{z_{k} \in \delta z} ( \hat{\sigma}^{k} _{22}\hat{\rho}\hat{\sigma}^{k} _{22}-\frac{1}{2} \hat{\sigma}^{k} _{22} \hat{\sigma}^{k} _{22} \hat{\rho}-\frac{1}{2}\hat{\rho}\hat{\sigma}^{k} _{22}\hat{\sigma}^{k} _{22} ).
\end{eqnarray}

This term only describes the situation where the atomic population remains in state $|1\rangle$ during the whole process.
If the pumping preparation is not optimum or if inelastic collisions are non-negligible, a population exchange term $\mathcal{L}^{coll}_{[1,2]}$ needs to be introduced. 

It is defined as 
\begin{eqnarray}
\mathcal{L}^{coll}_{[1,2]}[\hat{\rho}]=\gamma_{c} \sum_{z_{k} \in
\delta z}( \hat{\sigma}^{k} _{12}\hat{\rho}\hat{\sigma}^{k} _{21}-\frac{1}{2} \hat{\sigma}^{k} _{12} \hat{\sigma}^{k} _{21} \hat{\rho}-\frac{1}{2}\hat{\rho}\hat{\sigma}^{k} _{12}\hat{\sigma}^{k} _{21} )  \nonumber\\
+\gamma_{c} \sum_{z_{k} \in
\delta z}( \hat{\sigma}^{k} _{21}\hat{\rho}\hat{\sigma}^{k} _{12}-\frac{1}{2} \hat{\sigma}^{k} _{21} \hat{\sigma}^{k} _{12} \hat{\rho}-\frac{1}{2}\hat{\rho}\hat{\sigma}^{k} _{21}\hat{\sigma}^{k} _{12} ).
\end{eqnarray}

$\mathcal{L}^{coll}_{[1,2]}[\hat{\rho}]$ also affects the off-diagonal terms in the density matrix in the same way as $\mathcal{L}^{deph}_{[1,2]}[\hat{\rho}]$,  but as the sources of these two decoherence processes are different we monitor them separately.
It should be noted that this last term does not account for a pure loss of atoms out of the system, due to possible atomic motion out of the interaction region 
or atoms moving into other hyperfine states. 
We also assume the mean dephasing rates describing quantum jumps from $|1\rangle$ to $|2\rangle$ to be the same as the mean rates describing quantum jumps from $|2\rangle$ to $|1\rangle$ for simplicity.
 \begin{figure}[!ht]
\begin{center}
\includegraphics[width=\columnwidth]{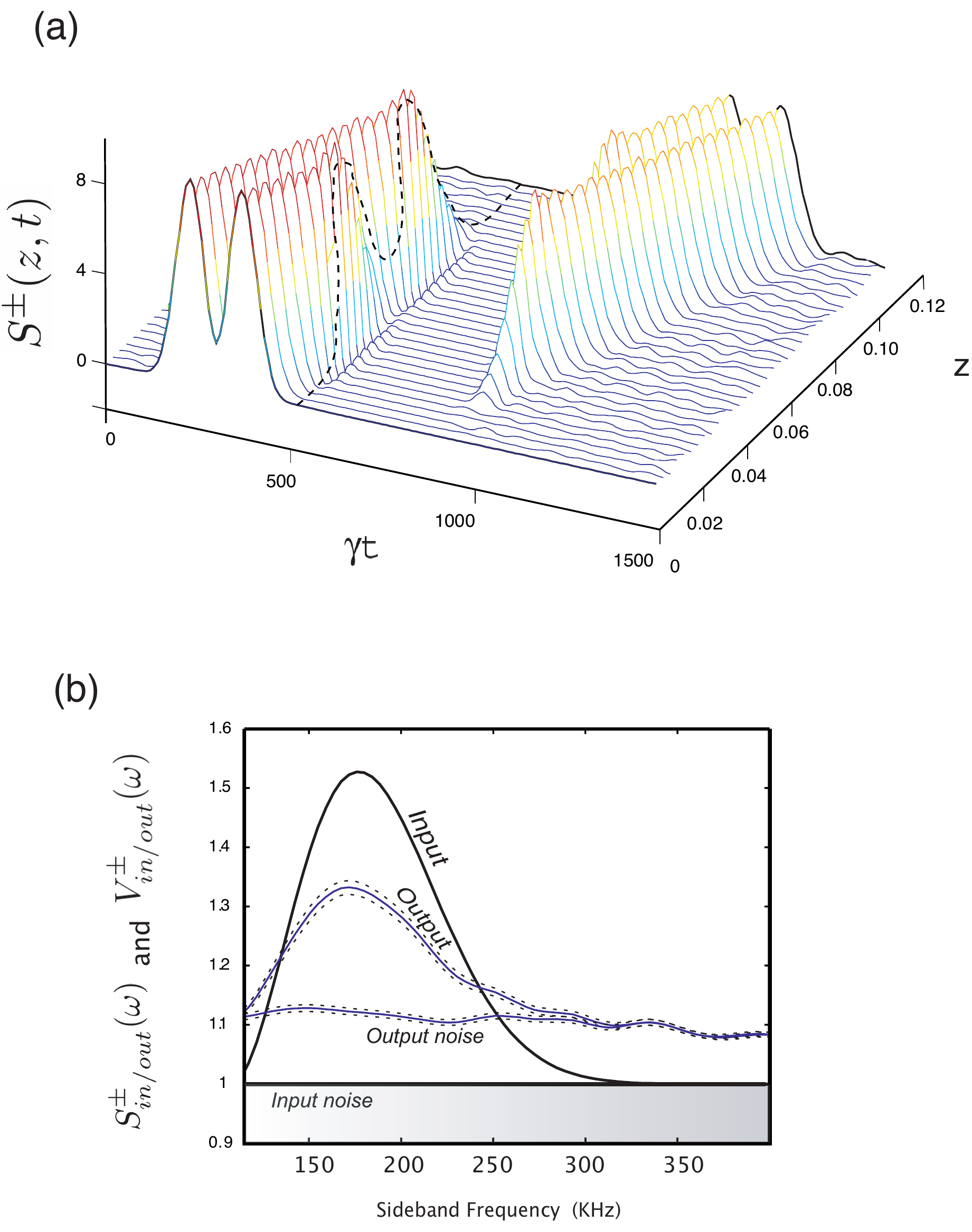}
\caption{Phase space numerical simulations of quantum information storage using EIT. 
Amplitude and phase modulations at 190 kHz are applied to the pulse. The decoherence rates are $\gamma_{0}=250$ Hz, $\gamma_{c}=100$ Hz.
a) 3D graph showing the storage of the probe amplitude quadrature on a time-space grid.
b) are the variances of the input/output fields for the amplitude and phase quadratures, 1 corresponds to the quantum noise limit. These simulations used 2000 trajectories.}
\label{stoch}
\end{center}
\end{figure}

To model this system, we used stochastic phase space methods, and worked with the Positive-P representation 
\cite{drummond}.
This phase space representation is computationally intensive but has the advantage of being exact as opposed to the truncated Wigner representation. 
We choose the following normal ordering of the operators 
\begin{equation}
(\hat{\mathcal{E}}^{\dagger},\hat{\sigma} _{13}^{\dagger},\hat{\sigma} _{23}^{\dagger},\hat{\sigma}_{12}^{\dagger},\hat{\sigma} _{33},\hat{\sigma} _{22},\hat{\sigma}_{11},\hat{\sigma}_{13},\hat{\sigma} _{23},\hat{\sigma} _{12},\hat{\mathcal{E}})
\end{equation}
and define 
\begin{equation}
\hat{\Xi}(\underline{\lambda},z)=\prod_{i} e^{\lambda_{i}\hat{O}_{i}(z)}
\end{equation}
where $\hat{O}_{i}(z)$ refers to the $i$th operator in our normally ordered definition and $\underline{\lambda}=(\lambda_{0}...\lambda_{i}...\lambda_{11})$ is a real vector.
The normally ordered characteristic function \cite{gardiner} is then
\begin{equation}
\chi(\underline{\lambda},z)={\rm Tr}(\hat{\rho}~\hat{\Xi}(\underline{\lambda},z)).
\end{equation}

The equations of motion for $\chi(\underline{\lambda},z)$ are calculated using the master equation (\ref{master}) and the commutation properties of the atomic and field operators. By taking the Fourier transform of the characteristic function equations of motion, and assuming a large number of atoms in each slice $\delta z$, 
a Fokker-Plank equation can be found. 
The Stratonovitch corrections are small compared to all the other variables and are not included in the SDE.
A set of nine c-numbers stochastic differential equations (SDE) describing the atomic dynamics with 18 uncorrelated noise terms arising 
from atomic fluctuations is then derived in the Ito form.
Their expressions are given in the Appendix A. The Maxwell equations for the probe envelope in a moving frame at the speed of light, are 
\begin{eqnarray}
\frac{\partial}{\partial z} \alpha(z,t) &=& \frac{igN}{c}\sigma_{3}(z,t)  \\
\frac{\partial}{\partial z} \beta(z,t)  &=& \frac{igN}{c}\sigma_{11}(z,t)
\end{eqnarray}
where the c-numbers $\alpha$ and $\beta$ represent the operators $\hat{\mathcal{E}}^{\dagger}$ and $\hat{\mathcal{E}}$, and $\sigma_{3}$, $\sigma_{11}$
correspond to the atomic operators $\hat{\sigma}_{13}$, $\hat{\sigma}_{13}^{\dagger}$.

The evolution of $\alpha$ and $\beta$ in space and time is computed when amplitude and phase modulations at a frequency $0.005\gamma$ are encoded onto a $50/\gamma$ long coherent input state. The envelope of the field then presents two cycles in both quadratures and allows the information to be encoded at frequencies where the pulse is shot noise limited.
We numerically evaluate the expectation values of the two quadrature operators
$\langle \hat{X}^{+}(z,t)\rangle=\overline{ \alpha(z,t)+\beta(z,t)}$ and $\langle \hat{X}^{-}(z,t) \rangle=-i~(\overline{ \alpha(z,t)-\beta(z,t)})$
and their noise spectrum $S^{\pm}(\omega) =\frac{c}{L}\overline{ X^{\pm}(z,\omega)X^{\pm}(z,-\omega)}$,
where the averaging is done over a large number of trajectories in phase space.
The noise floor $V^{\pm}(z,\omega)$ is obtained by turning off the signal on the probe.
We solved these stochastic equations using the numerical package XMDS \cite{xmds}
and chose parameters realistic to atom optics experiments with $^{87}$Rb atoms. 
The atomic density was chosen to be $10^{12}\,$cm$^{3}$ with a total length of 12 cm. 
At the moment the pulse is inside the medium, the coupling beam is switched off abruptly and turned back on $50/\gamma$ later. 
We chose for these particular simulations, a dephasing rate $\gamma_{0}=250$ Hz and an inelastic scattering rate $\gamma_{c}=100$ Hz.

Fig.~\ref{stoch} shows the results of this simulation where two quadratures of the multimode field have been stored in the presence of atomic noise.  
The stochastic simulations were averaged over 2000 trajectories.
Fig.~\ref{stoch} (a) shows the propagation of the amplitude quadrature of the modulated pulse through the cell. 
The results are identical for the phase quadrature and are not shown here.
We can see that the EIT-memory preserves the shape of the signal with minimal distortion.
To better quantify this we plot the power spectrum of the input and output fields in Fig.~\ref{stoch} (b). 
The asymmetry in the transmission reveals a frequency-dependent absorption of the pulse as it propagates through the system, characteristic of the EIT Lorentzian transmission window.
We also see that 60 $\%$ of the classical signal is absorbed and that extra noise is added to the field. Using the previously defined notation the transmission $\eta^{\pm}(\omega)=0.40$ 
and the excess noise $V^{\pm}_{\rm noise}(\omega)=0.12$. 
We will see in the last section if these conditions correspond to a quantum memory regime and describe the origin of the noise in the following section.

It should be noted here that an iterative procedure was recently proposed to optimize the coupling beam shape and power \cite{Gor}.
Here, we chose our (time independent) coupling beam Rabi frequency by maximizing the output signal without decoherence, i.e we found
a trade off between off line center absorption and the compression of the pulse required to fit the sample.
In this case the efficiency $\eta$ was found to be 80 \%, only limited by the lack of optical depth.
Using the same procedure, at higher densities and re-optimizing the coupling beam strength, we found the transmission to be close to unity.
Such time-bandwidth considerations are developed formally in next section B-2.

\subsection{Interpretation}

In this section we provide an explanation of the results found in the phase space simulations in the previous section.
We first discuss the effects of decoherences on the losses and atomic noise introduced during the light propagation.
We will show that excess noise can be understood as a preservation of the canonical commutation relations of the field in the presence 
of gain in the medium. We will quantify this by solving the Heisenberg-Langevin equations in the weak probe approximation in the case of information delay,
and compare it with a more general theory of amplification and attenuation.
We then describe the mapping and readout of the information encoded on the probe, derive boundaries for optimum storage, and quantify the maximum
information that can be stored in this system.
As in our numerical simulations, the process will be solved when the coupling beam is switched off abruptly.

\subsubsection{The role of decoherences}

We will here focus on the noise properties of the EIT as a delay line to explain the excess noise observed.
From the interaction Hamiltonian Eq. (\ref{hamiltonian}), we can obtain a set of Heisenberg-Langevin equations
\begin{eqnarray}
\dot{\hat{\sigma}}_{11} & = & \gamma \hat{\sigma}_{33} +
\gamma_{c} (\hat{\sigma}_{22} - \hat{\sigma}_{11}) - i g
\hat{\mathcal{E}} \hat{\sigma}_{31} + i g^{\ast}
\hat{\mathcal{E}}^{\dagger} \hat{\sigma}_{13} + \hat{F}_{11} \nonumber \\
\dot{\hat{\sigma}}_{22} & = & \gamma \hat{\sigma}_{33} +
\gamma_{c}(\hat{\sigma}_{11} - \hat{\sigma}_{22}) - i \Omega_c
\hat{\sigma}_{32} + i \Omega_c^{\ast} \hat{\sigma}_{23} +
\hat{F}_{22} \nonumber \\
\dot{\hat{\sigma}}_{13} & = & -(\gamma+\gamma_0/2+\gamma_c/2)\hat{\sigma}_{13} + i g
\hat{\mathcal{E}} (\hat{\sigma}_{11} - \hat{\sigma}_{33} )\nonumber \\
 &+& i\Omega_c \hat{\sigma}_{12} + \hat{F}_{13} \nonumber \\
\dot{\hat{\sigma}}_{32} & = &  -(\gamma+\gamma_0/2+\gamma_c/2) \hat{\sigma}_{32}+ i
\Omega_c^{\ast} (\hat{\sigma}_{33} - \hat{\sigma}_{22})\nonumber \\
 &-& i~g^{\ast} \hat{\mathcal{E}}^{\dagger} \hat{\sigma}_{12} +
\hat{F}_{32} \nonumber \\
\dot{\hat{\sigma}}_{12} & = & -(\gamma_{0}+\gamma_{c}) \hat{\sigma}_{12} - i g
\hat{\mathcal{E}} \hat{\sigma}_{32} + i \Omega_c^{\ast}
\hat{\sigma}_{13} + \hat{F}_{12} \nonumber \\
\frac{\partial}{\partial z}\hat{\mathcal{E}}&=& \frac{igN}{c} \hat{\sigma}_{13} \label{max}
\end{eqnarray}
where we have included the decays of the atomic dipole operators, 
the sources of decoherence introduced previously and their associated Langevin noise operators $\hat{F}_{ij}$ describing the coupling of the atoms to vacuum modes of large reservoirs.
The expressions for the Langevin correlations are calculated using the Einstein generalized equations \cite{Tannoudji,peng} and the non-zero contributions are given in Appendix (B). The system of equations ($\ref{max}$) will be solved to first order in $\hat{\mathcal{E}}$, $\gamma_c/\gamma$ and $\gamma_0/\gamma$.
To ensure efficient EIT, we will also assume $|\Omega_c|^2\gg(\gamma\gamma_0,\gamma\gamma_c)$. 

We first perform a steady state analysis of this system. 
Assuming the coupling beam Rabi frequency to be real, the atomic steady states are found to be

\begin{eqnarray}
\langle \hat{\sigma}_{11}\rangle &=& 1-2 \frac{\gamma_{c}}{\gamma}~,~\langle\hat{\sigma}_{22}\rangle = \frac{\gamma_{c}}{\gamma}~,~\langle\hat{\sigma}_{33}\rangle  = \frac{\gamma_{c}}{\gamma} \nonumber \\
\langle \hat{\sigma}_{12}\rangle &=& -\frac{g\langle\hat{\mathcal{E}}\rangle}{\Omega_c}~,~\langle \hat{\sigma}_{13}\rangle = \frac{ig\gamma_{0}}{\Omega_c^{2}} \langle\hat{\mathcal{E}}\rangle~,
~\langle \hat{\sigma}_{23}\rangle =  \frac{i\gamma_{c}}{\Omega_c}.\nonumber \\
\end{eqnarray}

We first note that the atoms are no longer fully pumped in the state $|1\rangle$ due to population exchange $\gamma_c$ and
a non-zero dipole $\langle \hat{\sigma}_{23}\rangle$ therefore appears on the coupling beam transition.
In this paper, however, we have assumed that the coupling beam is not depleted.
In order for these solutions to be consistent, we then need to find the regimes where the coupling beam is negligibly absorbed.
We do so by solving the following Maxwell equation for the coupling beam propagation 
\begin{equation}
\frac{\partial \Omega_c(z)}{\partial z}=\frac{ig^2N}{c} \langle \hat{\sigma}_{23}\rangle,
\end{equation}
the solution for which is 
\begin{equation}
\Omega^2_c(z)=\Omega^2_c(0)+2d\gamma\gamma_c z/L
\end{equation}
where $d=\frac{g^2NL}{\gamma c}$ is the optical depth of the medium.
Although the coupling beam intensity is absorbed linearly through the medium, a negligible depletion is guaranteed under the condition 
\begin{equation}
\frac{\Omega^2_c}{\gamma\gamma_c}\gg2d\label{pumpdepletion}
\end{equation}
which we will require in all the following calculations. We checked that this condition is verified in the above numerical analysis and the one presented in the last section.
We also note that because of the pure dephasing $\gamma_{0}$ a dipole $\langle\hat{\sigma}_{13}\rangle$ is created on the probe transition. 
A portion of the mean probe field is then absorbed by the medium by an amount $e^{-\alpha_{0}L}$, where
$\alpha_{0}=\frac{gN}{c}\frac{\gamma_{0}}{\Omega_c^{2}}$.

We will now calculate the evolution of the probe quantum field as it propagates through the medium in the same approximate regime.
To simplify the equations, the fast-decaying atomic variables will be adiabatically eliminated ($\frac{\partial}{\partial t}(\hat{\sigma}_{13},\hat{\sigma}_{23})  \ll \gamma$), 
making these equations only valid over timescales larger than the spontaneous emission time, which is the regime of interest for EIT. 
We follow the same procedure as in \cite{peng} and solve the equations in the Fourier domain. 
Using the steady state solutions listed above, we can eliminate the second order terms in the probe power and negligible Langevin noise contributions using Appendix B. 


The Maxwell equation for the field amplitude quadratures can be solved to give 
\begin{eqnarray}\label{fieldnoise}
\hat{\mathcal{E}}(z,\omega)&=&\hat{\mathcal{E}}(z,\omega)
e^{-\Lambda(\omega) z} \nonumber \\
-\frac{gN}{c} & \int_0^z& {\rm ds}~e^{-\Lambda(\omega)(z-s)}  \frac{\omega-i(\gamma_0+\gamma_c)}{\mu(\omega)}\hat{F}_{12}(s,\omega) \nonumber\\
+\frac{gN}{c} & \int_0^z& {\rm ds}~e^{-\Lambda(\omega)(z-s)}\frac{i\Omega_c}{\mu(\omega)}
\hat{F}_{13}(s,\omega) \label{quadratureout}
\end{eqnarray}
where $\mu(\omega)=\Omega_c^2-i\omega(\gamma+\gamma_d/2)$; $\gamma_d=\gamma_0+\gamma_c$ is the total decoherence rate
and the susceptibility of the medium is given by
\begin{equation}
\Lambda(\omega)=\frac{g^2N}{c}\frac{(\gamma_d-i\omega)(\langle\hat\sigma_{11}\rangle-\langle\hat\sigma_{33}\rangle)-i\langle\hat\sigma_{32}\rangle\Omega_c}{\mu(\omega)}. 
\end{equation}

%

The first part of Equation (\ref{fieldnoise}) describes the absorption
and phase shift of the probe propagating with a group velocity given by $v_{g}=-\omega/\Im(\Lambda(\omega))$ inside the EIT medium.
The last two terms in Equation (\ref{quadratureout}) correspond to atomic noise added to the field due to dephasing.

We now need to calculate the power spectrum of the output state as a function as the input state using Equations (\ref{powspect},\ref{fieldnoise}) 
and the Langevin correlations listed in Appendix B.
First we note that 
\begin{equation}
2\Re(\Lambda(\omega))=\frac{\Omega_c^2\langle[\hat{F}_{12},\hat{F}_{12}^{\dagger}]\rangle+\omega^2\langle[\hat{F}_{13},\hat{F}_{13}^{\dagger}]\rangle}{|\mu(\omega)|^2}
\end{equation}
links the linear absorption with the atomic noise, a direct consequence of the fluctuation dissipation theorem. 
This allows us to obtain
\begin{equation}\label{variance1}
S^{\pm}(z,\omega)=\eta(z,\omega)~S^{\pm}_{in}(\omega)+(1-\eta(z,\omega))(1+N_f)
\end{equation}
where $\eta(z,\omega)=e^{-2\Re(\Lambda(\omega))z}$, and
\begin{eqnarray}\label{Bnoise}
N_f&=&2\frac{\Omega_c^2\langle\hat{F}_{12}^{\dagger}\hat{F}_{12}\rangle+\omega^2\langle\hat{F}^{\dagger}_{13}\hat{F}_{13}\rangle}
{\Omega_c^2\langle[\hat{F}_{12},\hat{F}_{12}^{\dagger}]\rangle+\omega^2\langle[\hat{F}_{13},\hat{F}_{13}^{\dagger}]\rangle} \nonumber\\
&=&  \frac{4\gamma_c\Omega_c^2}{2\gamma_0\Omega_c^2+\omega^2(2\gamma+\gamma_0-3\gamma_c)}.
\end{eqnarray}

We note that the noise power spectrum is phase independent, indicating that the response of the medium is the same for both quadratures of the field. 
As expected, the normally ordered Langevin correlations are responsible for excess noise on the output field,
the strength of which depends on the ratio between normally ordered Langevin correlations and Langevin commutators.
However, because to first order in $\gamma_c/\gamma$, $\langle \hat{F}_{13}^{\dagger}\hat{F}_{13}\rangle=0$ (see Appendix B), spontaneous emission is not responsible for excess noise.
$\langle\hat{F}_{12}^{\dagger}\hat{F}_{12}\rangle=4\gamma_c$, so only the population shuffling term $\gamma_c$ transforms a initial coherent state into a mixed state.
Using the notations defined previously we have $V^{\pm}_{\rm noise}(\omega)=(1-\eta(\omega))N_f$.
The preservation of the commutation relations of the output field is ensured by the anti-normally ordered Langevin correlations that give the noise term $1-\eta$.
Indeed, When $\gamma_c=0$, we can check that we have have  $V^{\pm}(z,\omega)=1$.

In order to understand why population exchange between the ground states is responsible for noise, 
we will simplify the equations further and concentrate on sideband frequencies close to the carrier.
%

We first solve for the steady states with the only source of decoherence being the shuffling terms $\gamma_{c}(\hat{\sigma}_{11} - \hat{\sigma}_{22})$ and $\gamma_{c}(\hat{\sigma}_{22} - \hat{\sigma}_{11}$).
We find a new solution for the atomic polarisation $\langle \hat{\sigma}_{13}\rangle$ and insert this into the Maxwell equation to give 
$\langle\hat{\mathcal{E}}({\rm z})\rangle=\langle\hat{\mathcal{E}}_{\rm in}\rangle e^{az}$ where $a=\frac{gN}{c}\frac{\gamma_{c}}{\Omega_c^{2}}$.
This corresponds to a population exchange driven {\it amplification} of the probe field inside the medium, the energy for which will be provided by the coupling beam, up to a limit also set by Equation (\ref{pumpdepletion}).


This shuffling term alone is, however, not physically realistic.
As can be seen from the stochastic equation listed in the appendix and the Equations (\ref{max}), 
the Liouvillian $\mathcal{L}^{coll}_{[1,2]}$ also includes a ground state dephasing with mean rate $\gamma_c$ giving an extra linear loss $\alpha_c=\frac{gN}{c}\frac{\gamma_{c}}{\Omega_c ^{2}}$ similar to $\alpha_0$.
When solving for the steady state solutions in this case, with the same approximations as above, we find the net transmission close to zero frequency to be unity. 
The losses in fact exactly compensate for the gain, and the EIT medium no longer performs amplification.
Even though the transmission that includes $\mathcal{L}^{coll}_{[1,2]}$ does not depend explicitly on $\gamma_c$, this underlying
gain term results in the creation of excess noise.

Using Eq. (\ref{Bnoise}) close to $\omega=0$ and for $\gamma_0\neq0$ we can find the noise to be
\begin{eqnarray}\label{DCnoisefactor}
V_{{\rm noise}}&=&2\frac{\gamma_c}{\gamma_0}(1-e^{(a-\alpha)z});
\end{eqnarray}
whereas for $\gamma_0=0$,  $V_{\rm noise}=2 a z$.
 
A similar expression was found in the theory of two beam coupling developed in \cite{agarwal}.
The presence of excess noise on the output field was interpreted from the theory phase insensitive quantum amplifiers.
In \cite{caves} the signal to noise ratio of the optical field was shown to degrade in the presence of gain, 
and extra noise has to be inserted in the field equations to preserve the commutation relations. 
Precisely, it was shown that the output of an ideal linear amplifier with a gain factor $G>1$, relates to the input state by 
\begin{equation}
\hat{\mathcal{E}}_{\rm out}=\sqrt{G}~\hat{\mathcal{E}}_{\rm in}+\sqrt{G-1}~\hat{\mathcal{E}}^{\dagger}_{\nu}
\end{equation}
where $\hat{\mathcal{E}}^{\dagger}_{\nu}$ is a vacuum mode of the reservoir.
The power spectrum at the output of an ideal phase insensitive amplifier is then given by $S^{\pm}_{\rm out}=G S^{\pm}_{\rm in}+G-1$.

We will now follow the approach of Jeffers {\em et al}. \cite{jeffers} and will consider here a general theory for amplification and attenuation.
By concatenating $m$ amplifying and attenuating infinitesimal slices with linear amplification $1+a\delta z$ and attenuation $1-\alpha\delta z$, where $\delta z=z/m$ we will calculate the noise properties of the field at the output of such a sample and compare it to the previous result based on the Heisenberg-Langevin equations.
The power spectrum of the field at a slice $m$ is given by 
\begin{eqnarray}
S^{\pm}_m&=&(1+\frac{(a-\alpha)z}{m})^m (S^{\pm}_{in}-1)+1 \nonumber \\
&+&2 a \sum_{j=1}^m(1+\frac{(a-\alpha)z}{m})^{m-j}.
\end{eqnarray}

By going to the limit $m\rightarrow\infty$, converting the discrete slices into a continuous array, we get when $\alpha\neq a$
\begin{equation}\label{noisegain}
S^{\pm}(z)=\eta'(z)~S^{\pm}_{in}+(1-\eta'(z))(1+N_f')
\end{equation}
where 
\begin{equation}
N_f'=\frac{2a}{\alpha-a}~~{\rm and}~~\eta'(z)=e^{(a-\alpha)z}.
\end{equation}

This general treatment allows us to assess the amount of excess noise present at the output 
of a system when gain and attenuation are known quantities.
Eq. (\ref{DCnoisefactor}) can readily be found again by replacing $a$ and $\alpha$ by their value in the EIT system close to zero frequency which validates this interpretation.

We will now compare the signal to noise ratio and noise found by the present theory to the results given by the phase space treatment in the case of information delay.
The signal to noise ratio for both quadratures is defined by 
\begin{equation}
{\cal R}^{\pm}(z)=\frac{4 (\alpha^{\pm}(z))^{2}}{V^{\pm}(z) }.
 \end{equation}

\begin{figure}[!ht]
\begin{center}
\includegraphics[width=\columnwidth]{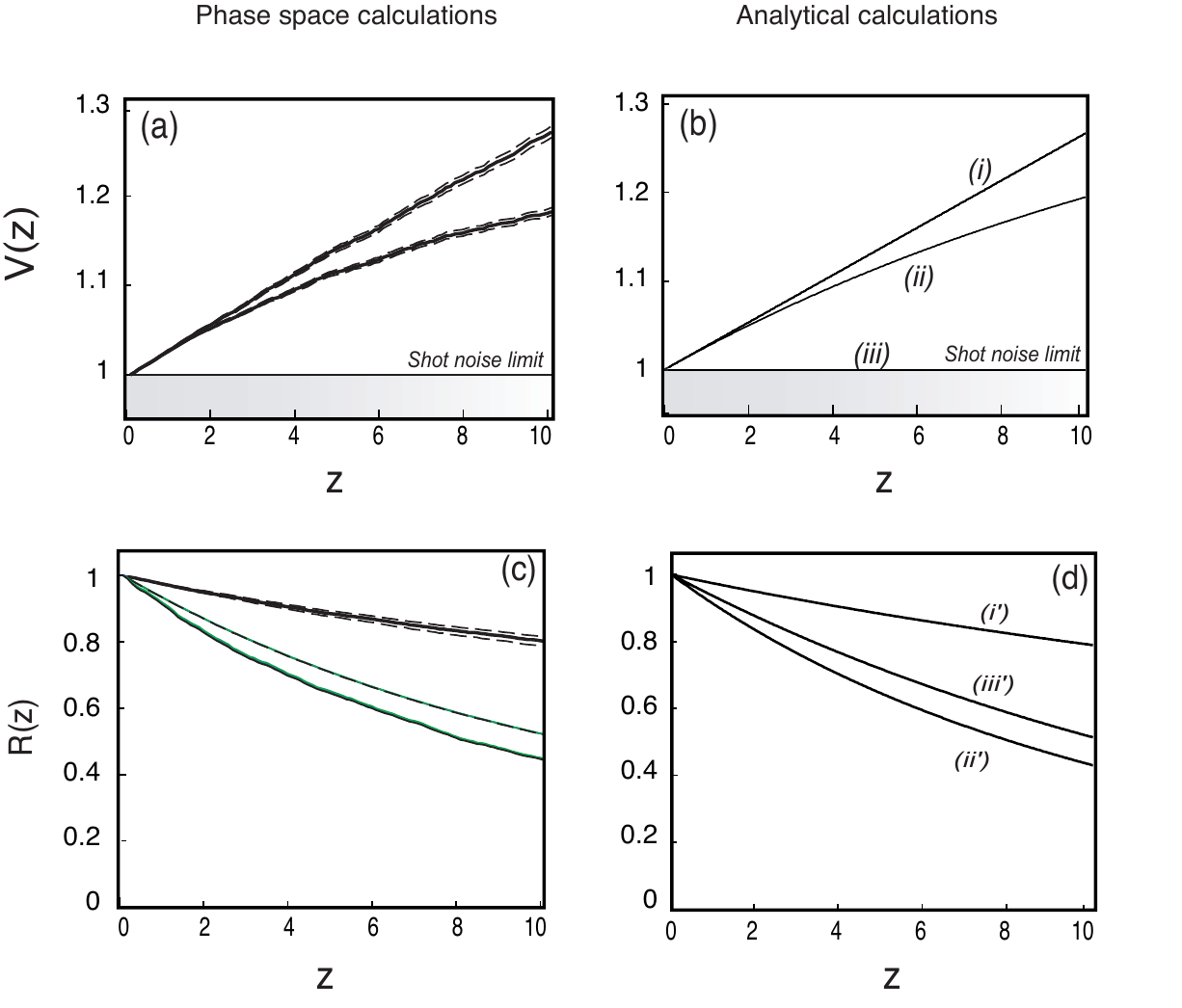}
\caption{Signal to noise ratios and noise results for the numerical simulations : (a) and (c), and analytical solutions : (b) and (d). (i) and (i') corresponds to $\gamma_c=0.005\gamma$ and $\gamma_0=0$; (ii) and (ii') corresponds to $\gamma_c=0.005\gamma$ and $\gamma_0=0.005\gamma$;   (iii) and (iii') corresponds to $\gamma_0=0.005\gamma$ and $\gamma_c=0$. The same parameters as for the phase space simulations of light storage were chosen.}
\label{noisedelay}
\end{center}
\end{figure}

Fig.~\ref{noisedelay} shows the evolution of the noise ($V(z)=1+V_{\rm noise}(z)$) and the signal to noise ratio as a function of the depth of propagation in three different 
situations. We consider the following decoherence combinations :  $(\gamma_0, \gamma_c) =(0,0.005\gamma)$, corresponding to curve (i); $(\gamma_0, \gamma_c)=(0.005\gamma,0.005\gamma)$, the curve (ii) and $(\gamma_0,\gamma_c) =(0.005\gamma,0)$, the curve (iii). Fig.~\ref{noisedelay}(a) and Fig.~\ref{noisedelay}(b) are the noise results from the two approaches and
Fig.~\ref{noisedelay}(c) and Fig.~\ref{noisedelay}(d) are the signal to noise ratio results.

Both theories are in good agreement. 
For curve (iii) there is no population exchange between the ground state and therefore the noise never exceeds the shot noise level.
For curve (i), the noise increases linearly as predicted when $\gamma_0=0$.
For curve (ii), the noise increases exponentially according to Equation (\ref{variance1}).

Fig.~\ref{noisedelay}(c) and Fig.~\ref{noisedelay}(d) compare the signal to noise ratios results from the two approaches and again a good agreement is found between the numerical simulations and the analytical solutions.
Even though the excess noise power is larger for (i) than for (ii) and (iii),  the signal transmission is 100 \% with $\gamma_c$ only, therefore the output signal to noise ratio is larger for (i') than for (ii') and (iii').

\
\subsubsection{Light storage}

In this section we present an analytical model of the light storage protocol.
Our treatment describes the transfer of information from the modulation sidebands 
of the probe beam to the collective atomic coherences in the sample and vice versa, 
taking into account the same  decoherence effects and the finite optical depth.
We will again consider fast switching and symmetric conditions for the writing and retrieval.


For lossless information delay the probe pulse does not need to be completely inside the sample.
However, light storage can only be performed efficiently if the group velocity is small 
enough such that the entire pulse is located inside the sample just before the switching of the coupling beam. In that case, information delay can be seen as light storage where the coupling beam
has been switched off and back on immediately afterwards.
Provided no noise is introduced due to the switching, which we verified numerically,
the excess noise on the output field can be 
determined by the previous information delay study and 
the Langevin terms will be ignored in the following.

The storage process is treated in three steps.
First, we describe the mapping of sidebands of the pulse of duration T to the 
atomic coherences in momentum space, this is the {\it writing stage}.
The second step, the {\it storage time} discusses
the influence of the decoherences when the coupling beam is off.
The last step, the {\it reading stage}, is the mapping of the information stored in momentum space back to a probe field.
We model the relaxation between the ground states with
the decoherence terms $\gamma_0$ and $\gamma_c$ introduced previously, in the same approximate regime. 
Similarly to \cite{matsko2}, to first order in $\hat{\mathcal{E}}$, two coupled linear equations can then be derived 
\begin{eqnarray} \label{init}
(\frac{\partial}{\partial z}+d'/L)~\hat{\mathcal{E}} (z,t)&=&\chi~\hat{\sigma}_{12}(z,t)\\\label{init2}
(\frac{\partial}{\partial t}+\Gamma_p)~\hat{\sigma}_{12}(z,t)&=&\nu~\hat{\mathcal{E}}(z,t)
\end{eqnarray}
%
where we introduce the quantities 
\begin{eqnarray}
\Gamma_p&=&\gamma_d+\frac{\Omega_{c}^{2}}{(\gamma+\gamma_d/2)}\nonumber\\
d'&=&d\frac{\gamma\langle\hat{\sigma}_{11}-\hat{\sigma}_{33}\rangle}{\gamma+\gamma_d/2}.
\end{eqnarray}
$\Gamma_p$ describes the pumping rate of photons from the coupling beam and $d'$ is the optical depth seen by the probe without coupling beam, in the presence of population shuffling. 
To simplify further the notations we also introduce
\begin{eqnarray}
\chi&=&-\frac{gN}{c}\frac{\Omega_c}{\gamma+\gamma_d/2}\nonumber\\
\nu&=&-ig\langle\hat{\sigma}_{32}\rangle-\frac{g\Omega_c\langle\hat{\sigma}_{11}-\hat{\sigma}_{33}\rangle}{\gamma+\gamma_d/2}.
\end{eqnarray}

Because these equations are linear, we will deal with the atomic and field as c-numbers and remove the hats on the operators.\\
\\
{\it Writing stage}\\


We define the {\it collective ground state coherence} as the Fourier transform in space of the locally averaged ground state coherence operator $\sigma_{12}(z,t)$,
\begin{eqnarray}
\sigma_{12}(k,t)=\frac{1}{L}\int_{0}^{L} \sigma_{12}(z,t) e^{i k z} dz.
\end{eqnarray}
We will see that under EIT conditions this quantity fluctuates with the same standard deviation as the input field.
During the writing stage the state of the probe at each point in space 
can be found using Eqs.~(\ref{init}) and (\ref{init2}) in the frequency domain.
As expected, the result is identical to the deterministic part of Equation (\ref{quadratureout}).
We can then obtain the mapping of the field in $\omega$ space to the coherences in momentum space 
when integrating Eq.~(\ref{init2}).
We consider the memory to work in the linearly dispersive regime, i.e the differential phase shift seen by all the spectral components of the field is the same.
This allows us to change variables from $\omega_0$  to $k_0 v_{g}$ when integrating Eq.~(\ref{init2}) and to get
\begin{eqnarray}\label{writing}
\sigma_{12}(k,t)&=&\int dk_{0}\mathcal{E}_{in}  ((k_{0}-k) v_{g}) \mathcal{D}_{W}(k_{0}, t)
\end{eqnarray}
where $\mathcal{D}_{W}$ is a ``distortion function" which quantifies the losses due to the finite bandwidth $\zeta(\omega)=\Re(\Lambda(\omega))$, and the finite length of the cell. 
It is given by
\begin{eqnarray}
\mathcal{D}_{W}(k_{0}, t)&=&\frac{\nu v_{g}}{\Delta \omega}\Big( \frac{ e^{ (ik_{0}-\zeta((k-k_{0})v_{g}  )   )L}-1  }{   ik_{0}L-\zeta((k-k_{0})v_{g})L  }\Big)\nonumber\\
& & \times \Big( \frac{ e^{(\Gamma_p-i(k-k_{0})v_{g}) t}-1  }{ \Gamma_p-i(k-k_{0})v_{g} }\Big).
\end{eqnarray}
The integration of Eq.~(\ref{writing}) is performed between $k-\Delta \omega / (2v_{g})$ and $ k+\Delta \omega / (2v_{g})$ where $\Delta \omega=1/T$.

We now require the medium to be optically thick,  $d'\gg kL$, and the frequency where the information is encoded 
to be smaller than the pumping rate $\Gamma_p$.
Both these conditions ensure a high efficiency of the writing process as we will see.
In this regime Eq.~(\ref{writing}) reduces to 
\begin{eqnarray}\label{downsampling}
\sigma_{12}(k,t) &=& \frac{\nu v_{g}}{\Delta\omega\Gamma_p}(1-e^{-\Gamma_p t}) \nonumber\\
& & \times \int  dk_{0}{\mathcal{E}}_{in} ((k_{0}-k)v_{g}) \rm{sinc}(  \frac{k_{0}L}{2} ).
\end{eqnarray}

This equation describes a downsampling of the information from the probe field to the atoms due to a finite optical depth during its compression. 
The information is loaded at a rate $\Gamma_p$ onto the collective ground state coherences.
This process is much faster than the time it takes for the pulse to enter the sample (which is on the order of $T$).
When Fourier transforming back to the spatial coordinate at a time $t_{\rm off}\simeq T$, we get an expression in the form of a convolution
\begin{eqnarray}\label{zspace}
\sigma^{t_{\rm off}}_{12}(z) &=& \frac{\nu}{\Gamma_p} {\rm sinc}(\Delta k z)*\big[H(L)\mathcal{E}_{\rm in} (-z/v_{g})\big]
\end{eqnarray}
where $H(L)$ is a top hat function defining the atomic sample boundaries. 
For the probe pulse to fit the atomic sample we then require the duration of the pulse to satisfy the relation $T\ll L/v_{g}$.
In this case, and provided the spatial components of the probe are well within the sample ($\Delta k z\ll1$), there is no loss of information and Eq. (\ref{zspace}) can be written
\begin{equation}\label{store}
\sigma_{12}^{t_{\rm off}}(z) = -\frac{\nu}{\Gamma_p} \mathcal{E}_{\rm in} (-z/v_{g}). \\ \nonumber
\end{equation}
In this process the statistics of the probe field is distributed onto the atomic ground state as it propagates through the medium.
Most of its energy is transferred to the coupling beam and has left the cell at the speed of light.
\\\

{\it Storage time}\\
 
The coupling beam is now switched off at $t=t_{\rm off}$. 
The evolution of the atomic coherence and of the remaining probe inside the medium will again be solved in the adiabatic limit.
The system of equations describing this process is 
\begin{eqnarray}
 0 & = & -(\gamma+\gamma_d/2) \sigma_{13} + i g
\mathcal{E} (\sigma_{11} - \sigma_{33} ) \nonumber \\
 0 & = &- (\gamma+\gamma_d/2)\sigma_{32}- i
g^{\ast} \mathcal{E}^{\dagger} \sigma_{12} \nonumber \\
\frac{\partial}{\partial t} \sigma_{12} & = & -\gamma_d \sigma_{12} - i g
\mathcal{E} \sigma_{32}. \nonumber \\
\end{eqnarray}

The Maxwell equation describing the time evolution of the spatial modes of the remaining probe inside the cell is then
\begin{equation}
(\frac{\partial}{\partial t} +ikc+1/\tau)\mathcal{E}(k,t)=0
\end{equation}
where $1/\tau=\frac{g^2N}{\gamma}(\langle\sigma_{11}\rangle-\langle\sigma_{33}\rangle)$.
After a time $3\tau$ the majority of the probe energy is then absorbed by the medium. 
The ground state coherences depend on the probe field to second order so this short process does not affect
the stored information.

Assuming that the coupling beam is switched back on at a time $t=t_{\rm on}$, we integrate Eq.~(\ref{init2}) from $t=t_{\rm off}$ to $t=t_{\rm on}$ to obtain
\begin{eqnarray}\label{waiting}
\sigma_{12}^{t_{\rm on}}(k)&=& e^{-\gamma_{d} (t_{\rm on}-t_{\rm off})} \sigma^{t_{\rm off}}_{12} (k)
\end{eqnarray}
which describes a simple exponential decay of the coherences over time due to a non-zero dephasing rate $\gamma_{d}$.\\\

{\it Reading stage}\\

To describe the reading stage we evaluate the coherences in the presence of a field on the probe transition in momentum space.
We first solve for $\sigma_{12}(k,t)$ independently of the probe field by combining Eq.~(\ref{init}) and Eq.~(\ref{init2}). We obtain
\begin{equation}
\sigma_{12}(k,t)=e^{-\beta(k) t} \sigma^{t_{on}}_{12} (k)
\end{equation}
where 
\begin{equation}
\beta(k)=\Gamma_p-\frac{\nu\chi}{d/L-ik}.
\end{equation}

We then follow the same procedure as in the writing stage.
The Maxwell equation (\ref{init}) is solved in $\omega$ space to give

\begin{eqnarray}\label{reading}
\mathcal{E} (z,\omega)&=& \int dk_{0} ~\sigma^{t_{\rm on}}_{12}(\frac{\omega-\omega_{0}}{v_{g}}) \mathcal{D}_{R}(z,\omega_{0})
\end{eqnarray}
where $\mathcal{D}_{R}$ is a distortion function now affecting the transfer from the atomic coherences to the field and is given by
\begin{eqnarray}
\mathcal{D}_{R}(z,\omega_{0})&=&\frac{\chi}{\Delta k v_{g} } 
\Big(\frac{ e^{     (i\omega_{0}-\zeta'(\frac{\omega-\omega_{0}} {v_{g} } )  )t   }  -1  }{    (i\omega_{0}T-\zeta'(\frac{\omega-\omega_{0}}  {v_{g}})  )T   }\Big)\nonumber\\
& & \times \Big( \frac{ e^{(d'/L-i\frac{\omega-\omega_{0}}{v_{g}} ) z}-1  }{d'-i\frac{\omega-\omega_{0}}{~v_{g}}L  }\Big)
\end{eqnarray}
where $\zeta'(k)=\Re(\beta(k))$.
The integration of Eq.~(\ref{reading}) is performed between $\omega-v_{g}\Delta k / 2$ and $\omega+v_{g}\Delta k / 2$.
Under the same condition as for the writing stage (high density and small enough pumping rate $\gamma_p$) we obtain 
\begin{eqnarray}\label{ret}
\mathcal{E} (z,\omega)&=& \frac{\chi L}{v_g\Delta k d'} (1-e^{-d' z/L}) \nonumber\\
& & \times \int dk_{0} \sigma^{t_{\rm on}}_{12} (\frac{\omega-\omega_0}{v_{g}}) \, {\rm sinc}(\frac{\omega_0 T}{2}).
\end{eqnarray}

The downsampling also occurs when the information is transferred from the ground state coherences to the probe due to the finite optical depth.
We can again transform this expression in time and space to obtain the field at the output of the sample 
\begin{eqnarray}\label{downsamplingreading}
\mathcal{E} (z,t)&=& \frac{\chi L}{d'}\rm{sinc}(\Delta \omega t)*\big[H(T)\sigma^{t_{on}}_{12}(-v_{g}t)\big]
\end{eqnarray}

This expression can be simplified in the case where the compressed probe pulse
 fits entirely within the atomic sample, i.e when the duration of the pulse $T$  satisfies the relation $T\ll L/v_{g}$.
There is then no loss of information and Eq. (\ref{downsamplingreading}) can be written 
\begin{eqnarray}\label{ret2}
\mathcal{E} (z,t)&=& \frac{\chi L}{d'}\sigma^{t_{on}}_{12}(-v_{g}t)
\end{eqnarray}

Using continuity arguments one can combine Eq.~(\ref{ret2}), Eq.~(\ref{store}) and Eq.~(\ref{waiting}) one can then see that 
\begin{eqnarray}\label{final}
\mathcal{E}_{\rm out}&=&\frac{\nu \chi L}{\Gamma_p d'}e^{-\gamma_d (t_{\rm on}-t_{\rm off})}\mathcal{E}_{\rm in}
\end{eqnarray}
This expression relates the input and output states in the presence of pure dephasing and population exchange between the ground states. One can show that when $\gamma_d=0$, the output is the perfect replica of the input state.

The result given by Eq. (\ref{final}) was obtained after making two main assumptions, $L\gg v_{g}T$ and $\Delta\omega\gg\Gamma_p$, 
which can be gathered in this inequality relation
\begin{eqnarray} \label{inequality}
v_g/L\ll\Delta\omega\ll\Gamma_p
\end{eqnarray}

The information has to be encoded at frequencies that satisfy this relation for perfect storage efficiency.
The lower bound has to be satisfied for the pulse to fit the atomic sample. With a long input pulse, i.e a small spectral extent $\Delta \omega$, a high density
or a weak coupling beam is required, whereas for a short input pulse $v_g/L$ can be made larger.
On the other hand, the upper bound defines the minimum EIT bandwidth tolerable to minimize the losses. 
A short input pulse will require a large coupling beam power, whereas 
a weaker coupling beam power (narrower EIT bandwidth) can be used.
The time-bandwidth product of the system ($ \Gamma_p \times v_g/L$) can in fact be found to be $d'$, 
i.e the number of independent samples from the probe that can be faithfully stored depends only on the density. 
Not only does $\gamma_c$ introduce excess noise on the output probe mode, but it also reduces the time bandwidth product at a given density.
At infinite density one can than store an infinitely broad probe spectrum.

This result is similar to the one found in the general case of time varying coupling beam in \cite{Gor}.
With a coupling power calculated by equating the length of the pulse in the cell with the length of the cell,
$\Omega_c^2=d\gamma/T$, the condition $T d \gamma\gg 1$ could be found. 
Our condition however, requires the full width at half maximum of the pulse to be smaller than the cell length.
As also mentioned in \cite{Gor}, the input and output pulses durations (which are identical in our case) also have to satisfy the relation $T \gamma_d\ll1$ for the
information to be imprinted onto the atoms before the pulse is absorbed.
We note that provided Eq. (\ref{inequality}) is verified, the condition $\Omega^2  / d\gamma  \ll \gamma_d$ is a sufficient condition for $\gamma_d T \ll 1$ to hold.
This is the case in all the numerical simulations presented in this paper.



\section{Quantum Information Benchmarks}

In this section, we will investigate the storage of optical information from a quantum informatic perspective.  We will benchmark the results obtained in our modeling against known quantum information criteria.  These criteria, which are fidelity, signal transfer coefficients and conditional variances,  will enable the determination of whether a quantum strategy has been used in the storage and readout of a quantum state; whether an EIT based quantum memory is possible in an experimentally realistic situation, and whether the output of the storage process is indeed the best clone of its input.

\begin{figure}[!ht]
\begin{center}
\includegraphics[width=\columnwidth]{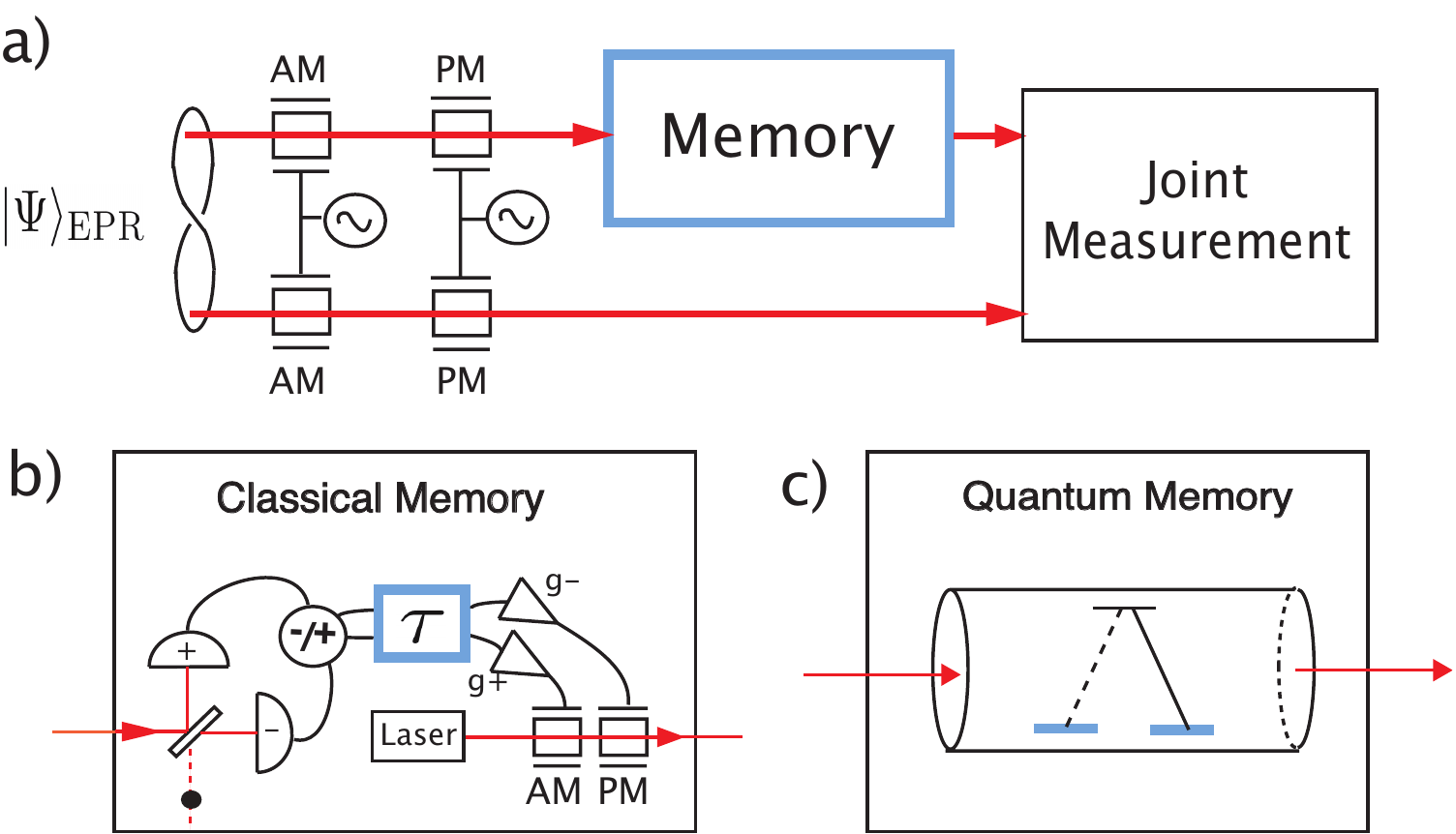}
\caption{(a) General schematics for characterizing an optical memory.  A pair of EPR entangled beams are encoded with amplitude and phase quadrature information.  One of these beams is injected into, stored and readout from the optical memory whilst the other is being propagated in free space.  A joint measurement with appropriate delay is then used to measure the quantum correlations between the quadratures of the two beams.  (b) A classical teleporter scheme used as an optical memory.  The input state is measured jointly on both quadratures using two homodyne detection schemes.  Analogous to classical teleportation the measured information is stored for time $\tau$ before fed-forward onto an independent laser beam with a feedforward gain, $g$.  The feedforward gain is analogous to a transmission of $\sqrt{\eta(\omega)}$ for EIT based memories.  (c) Quantum memory using EIT.  The input state is stored in the long lived ground state coherence of three level atoms in $\Lambda$ configuration.}
\label{setup}
\end{center}
\end{figure}

Fig.~\ref{setup} shows the schematics of our quantum memory benchmark.  It was shown in Ref.
\cite{cirac} that the optimal classical measure and prepare strategy for optical memory is the classical teleporter scheme as shown in Fig.~\ref{setup} (b), and we therefore benchmark the performance of our EIT quantum memory against this setup.  In this classical scheme, the storage time can be arbitrarily long without additional degradation.  However, two conjugate observables cannot be simultaneously measured and stored without paying a quantum of duty 
\cite{Braunstein,ralph}.  
Moreover, the encoding of information onto an independent beam using amplitude and phase modulators will also introduce another quantum of noise.  In total, the entire process will incur an additional two quanta of noise onto the output optical state.  

Possibly the best known benchmark in quantum information protocols is the fidelity which measures the wave function overlap between the output and input states. It is given by

\begin{equation}
\mathcal{F} =\langle\Psi_{\rm in} | \hat{\rho}_{\rm out} | \Psi_{\rm in}\rangle.
\end{equation}

which, in the Wigner representation can be written

\begin{equation}\label{fideli}
\mathcal{F} = 2\pi\int \!\!\! \int {\rm W}_{\rm in}({\rm X}^{+},{\rm X}^{-}){\rm W}_{{\rm out}}({\rm X}^{+},{\rm X}^{-}){\rm dX}^{+}{\rm dX}^{-}.
\end{equation}

For Gaussian states with coherent amplitude $\alpha^{\pm}$ and power spectrum $S^{\pm}$, the Wigner functions is 

\begin{equation}\label{Wig}
{\rm W}({\rm X}^{+},{\rm X}^{-})=\frac{2}{\pi S^{+}S^{-}}e^{-\frac{({\rm X^{+}}-2\alpha^{+})^2}{2S^{+}}-\frac{({\rm S^{-}}-2\alpha^{-})^2}{2S^{-}}}
\end{equation}

The fidelity of the classical teleporter scheme can be easily calculated \cite{bowen} using formula (\ref{Wig},\ref{fideli}) and gives  

\begin{equation}\label{fidelityeq}
  \mathcal{F}=  \frac{2~e^{-k^{+}-k^{-} }  } {\sqrt{(2+V^{+}_{\rm noise})(2+V^{-}_{\rm noise})}}
\end{equation}
where $k^{\pm}=\alpha^{\pm}_{\rm in}~ (1-g^{\pm})^{2} /(2+V^{\pm}_{\rm noise})$, $V^{\pm}_{\rm noise}$ are the noise variances of the output field for the amplitude and phase quadratures and $g^{\pm}$ is the feedforward gain.  For an ideal classical memory with unity gain, $g^{\pm}=1$, and one can see from the previous argument that a coherent input state will give $V^{\pm}_{\rm noise}=2$, thus giving a {\it classical limit} of $\mathcal{F} \ge 0.5$.  It has been shown by Grosshans and Grangier 
\cite{grosshans2} that when the fidelity of a teleporter $\mathcal{F} \ge 2/3$, the output state is guaranteed to be the best cloned copy of the input state.  This fidelity limit called the {\it no-cloning limit} for teleportation corresponds to the addition of only one quantum of noise in the entire process.

\begin{figure}[!ht]
\begin{center}
\includegraphics[width=\columnwidth]{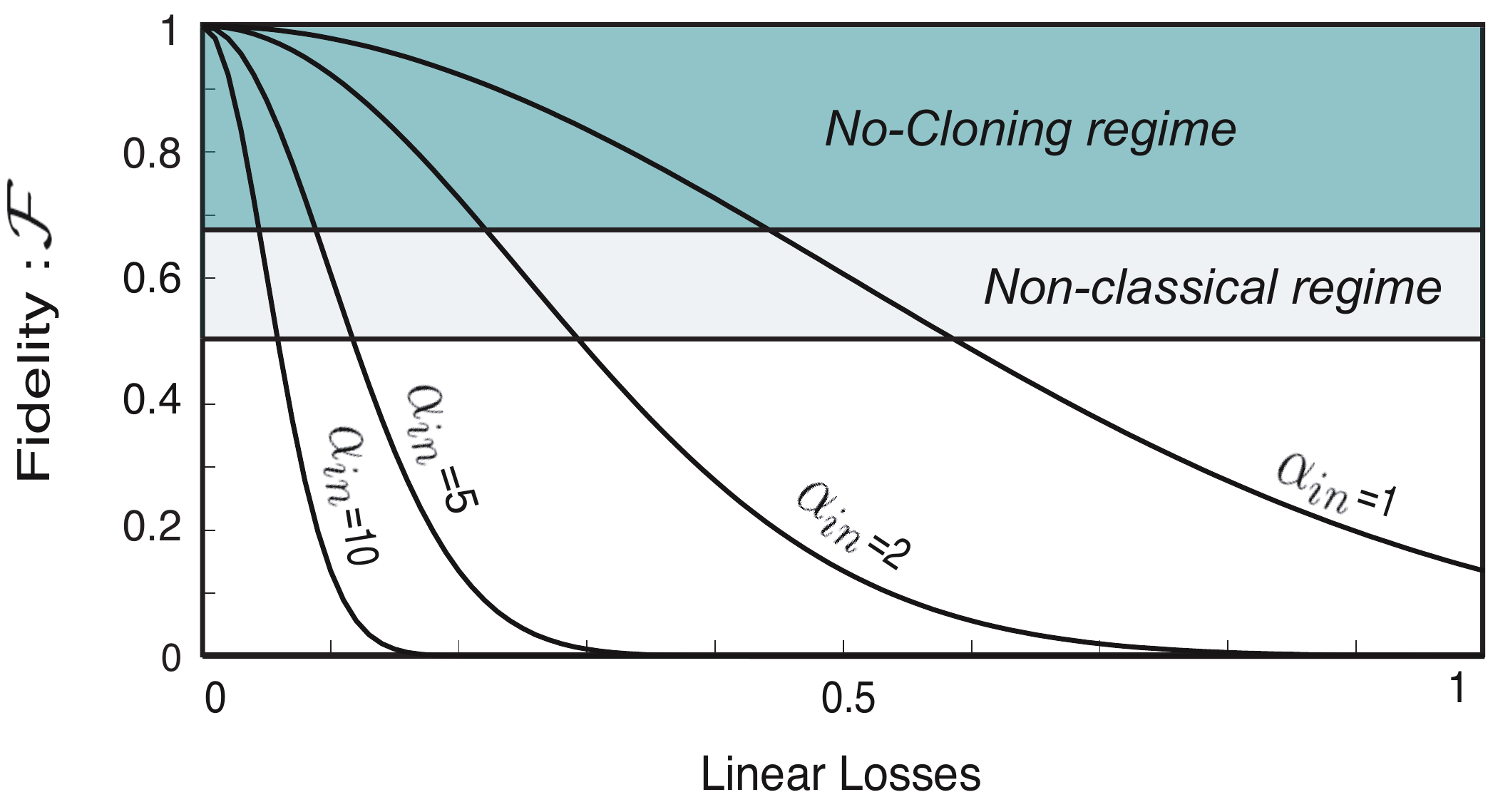}
\caption{Fidelity as a function of memory loss for EIT for $\alpha_{\rm in}(\omega)=10,5,2,1$.  The non-classical and the no-cloning regimes are reached when $\mathcal{F} \ge 1/2 $ and $\mathcal{F} \ge 2/3$, respectively.}
\label{fidelity}
\end{center}
\end{figure}

The use of entanglement in the context of quantum teleportation, or for example EIT for 
quantum memories is necessary to break these limits. We now quantify EIT-based quantum memories using this criterion. There is a direct analogy
between the feedforward gain, $g$ and the EIT transmission $\sqrt{\eta(\omega)}$.
Fig.~\ref{fidelity} shows the behavior of $\mathcal{F}$, as defined in Equation (\ref{fidelityeq}) with $g=\sqrt{\eta(\omega)}$, with varying memory loss for different coherent state amplitudes. 
We note that the maximum amount of memory loss tolerable for beating both limits are dependent on the coherent amplitudes of the input states.  
This shows that fidelity is a state dependent measure.  

The formula for the fidelity can be extended to mixed input states using
\begin{equation}
 \mathcal{F}=\big[ \rm tr(\sqrt{    \sqrt{\hat{\rho}_{\rm in}} \hat{\rho}_{\rm out}   \sqrt{\hat{\rho}_{\rm in}}   }) \big] ^{2}.
 \end{equation}
Jeong {\it et al.}  \cite{jeong} showed that this formula can again be used to benchmark quantum information protocols.  Nevertheless, characterizing quantum memory using the state dependent fidelity as a measure will be complicated for exotic mixed states.

 \begin{figure}[!ht]
\begin{center}
\includegraphics[width=\columnwidth]{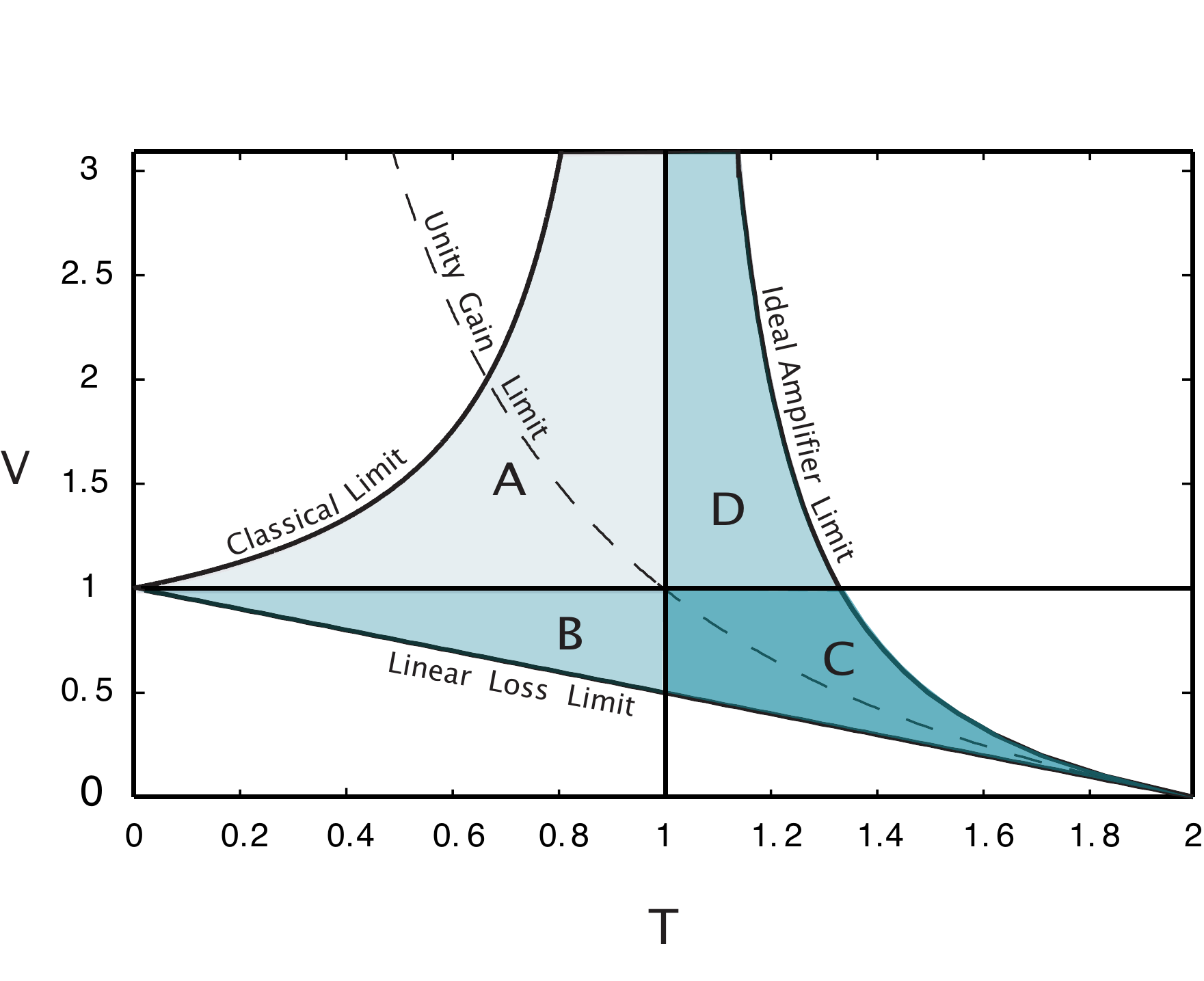}
\caption{Diagram for total signal transfer coefficient T versus conditional variance product V.  The classical limit line shows the optimal performance of a classical teleporter.  The passive loss limit defines the performance of an EIT-based memory that does not produce excess noise. 
The unity gain curve is obtained with increasing excess noise in an EIT system with no loss. 
The Amplification limit corresponds to the ideal amplifier limit.
Regions A, B, C and D correspond to the quantum regime; Regions B and C represent the regime where EPR entanglement is preserved; Region D is the lossless amplification region; Region C denotes the no-cloning limit.}
\label{TV}
\end{center}
\end{figure}

An alternative measure to fidelity for the characterization of quantum information protocols was proposed by Grangier {\it et al.} 
\cite{Grangier} for quantum non-demolition measurement and by Ralph and Lam 
\cite{ralph} for quantum teleportation.  This alternative uses the signal transfer coefficients, $T^{\pm}$, and the input-output conditional variances, $V_{\rm cv}^{\pm}$ to establish the efficacy of a process. The conditional variances and signal transfer coefficients are defined as
\begin{eqnarray}
V_{\rm cv}^{\pm}&=&V^{\pm}_{\rm out}-\frac{|\langle \hat{X}^{\pm}_{\rm in}\hat{X}^{\pm}_{\rm out}\rangle|^{2}}{V^{\pm}_{\rm in}}\\
T^{\pm}&=&\frac{{\cal R}^{\pm}_{\rm out}}{{\cal R}^{\pm}_{\rm in}}
 \end{eqnarray}
where ${\cal R}^{\pm}_{\rm out/in}$ is the signal to noise ratio of the output/input field defined by 
\begin{equation}
{\cal R}^{\pm}_{\rm in/out}=\frac{4 (\alpha^{\pm}_{\rm in/out})^{2}}{V^{\pm}_{\rm in/out} }.
 \end{equation}
We now define two parameters that take into account the performances of the system on both conjugate observables 
\begin{eqnarray}
V & = & \sqrt{V_{\rm cv}^{+}V_{\rm cv}^{-}}, \\
T & = & T^{+}+T^{-}.
 \end{eqnarray}
 
Fig.~\ref{TV} shows the plot of a TV-diagram.  Similar to the fidelity, there are corresponding classical and no-cloning limits in the TV-diagram for a teleporter or an optical memory.  It can be shown that a classical teleporter cannot overcome the $T>1$ or $V<1$ limits.  By tuning the feedfoward gain, $g$, a classical teleporter will perform at best at the ``classical limit" curve as shown in Fig.~\ref{TV}.  
Ref~\cite{ralph} shows that this classical limit can be surpassed using quantum resource (Region A).  With limited quantum resource, it is possible to have an output state with $V < 1 $ (Region B).  When the input state is from a pair of entangled beams, this performance corresponds to the preservation of EPR entanglement at the output 
\cite{reid}.  With a stronger quantum resource,  $T>1$ and $V<1$ can be satisfied simultaneously.  Grosshans and Grangier 
\cite{grosshans2} showed that under these conditions the output state represents the best cloned copy of the input.  The lower right quadrant of the TV-diagram (Region C) therefore corresponds to  the no-cloning regime.

We now characterize the EIT-based quantum memory in terms of the TV diagram.  When an EIT system does not generate excess noise, the performance of the memory is described by the linear loss limit line.  Assuming that the transmission through the EIT medium is identical for both quadratures, it can be shown that $V  =  1-\eta(\omega)$ and $T = 2 \eta(\omega)$.  We note that the result suggests an EIT with linear loss will surpass the classical limit independent of $\sqrt{\eta(\omega)}$.  This is because unlike the classical teleporter, the output state obtained from a linear loss EIT is not being measured throughout the transmission.  Thus there is no measurement quantum duty for all transmittivities.  Moreover, an input entangled state through a linear loss device will always preserve some entanglement at the output.  

However, when some excess noise is introduced in the storage process, $T$ will decrease and $V$ increase more rapidly.
This is the case if for example some amplification is involved. 
Indeed we have seen in the previous section that $G-1$ quanta of noise will be introduced for a lossless memory with a gain G in order to preserve the commutation relations at the output.
The performance of the memory is then described by the ideal lossless amplifier line on the TV diagram where one can show that $V=G-1$ and $T=2 G/ (2G-1)$. 
The optimum situation will then be when the gain of the amplifier is unity, so that $T=2$ and $V=0$. 
As the gain increases the memory no longer performs in the no-cloning regime and reaches region (D) where no quantum correlation exists between the input and output states anymore,
but the signal transfer is always larger than what a classical memory could achieve.

There are other possible sources of noise that do not amplify the signal.
For example, any transfer between the coupling beam and the probe via non-linear processes \cite{matsko2} or non-ideal polarizers will contribute to excess noise.  We can introduce the excess noise phenomenologically with $V_{\rm noise}$, which we can assume to be quadrature independent.  
The TV performance is now given by $V=1-\eta(\omega)+V_{\rm noise}$, and $T=2~\eta(\omega)/(1+V_{\rm noise})$.  
Unlike classical teleportation and in the absence of an amplification process, the EIT gain $\sqrt{\eta(\omega)}$ can only be less than or equal to unity.  
If we assume perfect transmittivity with $\sqrt{\eta(\omega)} = 1$, increasing excess noise produces the unity gain curve in the TV diagram.  We note that although input signal is perfectly transmitted, the excess noise leads to a degradation on both $T$ and $V$. 

\begin{figure}[!ht]
\begin{center}
\includegraphics[width=8cm]{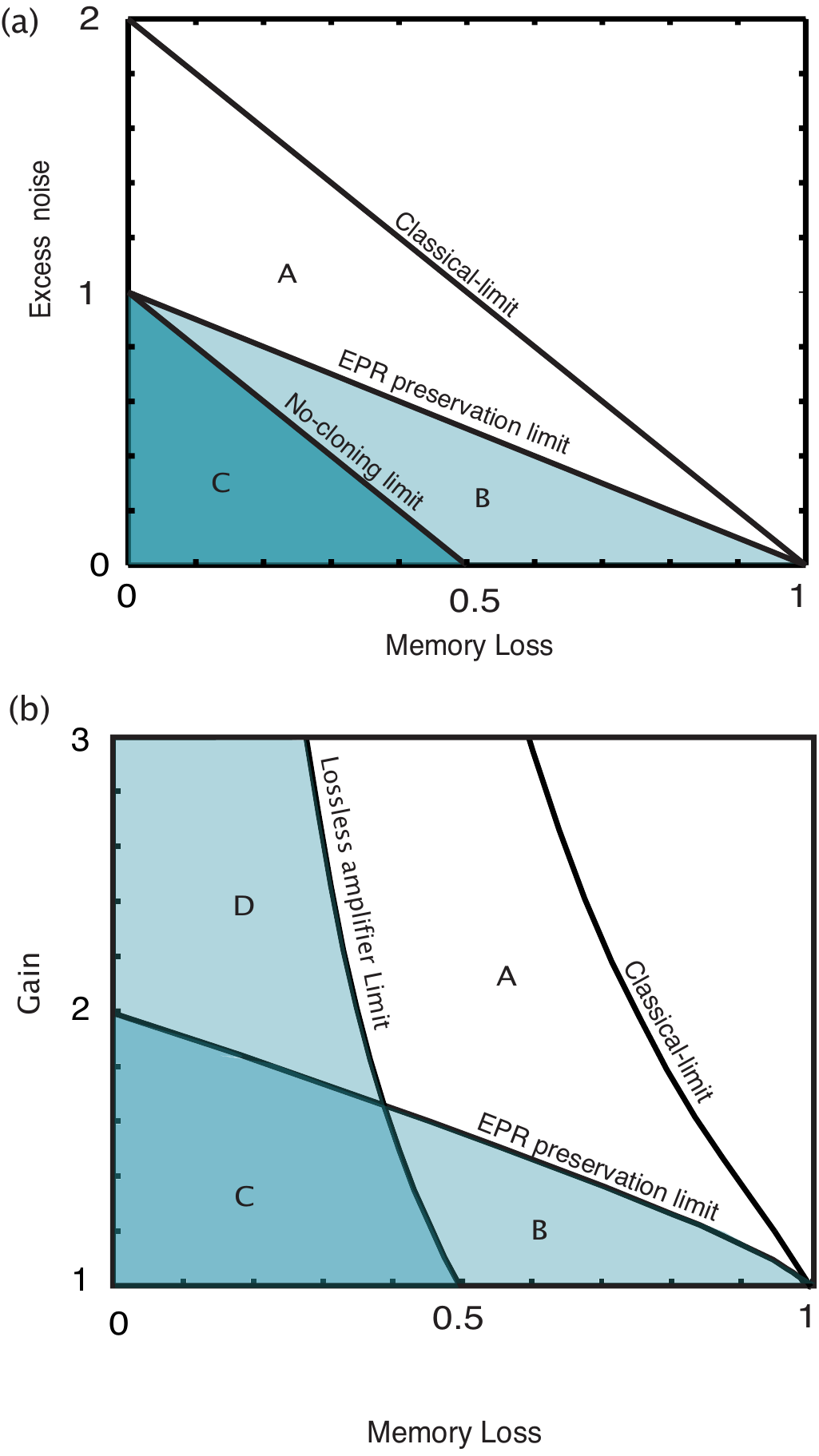}
\caption{Classical, EPR and no-cloning regimes plotted as a function of EIT linear loss and excess noise in (a) . 
A is the non-classical regime; B is the EPR regime and C is the no-cloning regime.
These limits are drawn in fig (b) on a loss-gain plot and have been derived from Equation (\ref{noisegain}). Contrary to the case of (a), with a large enough gain and sufficiently low losses in the memory, Region (D) can be reached.}
\label{Lossnoise}
\end{center}
\end{figure}

We will now define the parameters required to reach the quantum regime in both cases.
We plot these quantum regimes with excess noise versus linear loss in Fig.~\ref{Lossnoise}(a), and gain versus loss in Fig.~\ref{Lossnoise}(b).
In particular, to define Fig.~\ref{Lossnoise}(b) we calculated the $T$ and $V$ corresponding to the situation described by Equation (\ref{noisegain}).
Then we found the gain and losses for which the EIT-performance crosses our benchmarks.
These diagrams determine whether an experiment is sufficiently low noise and transmissive for quantum information storage.  
The no-cloning limit can only be surpassed when $\sqrt{\eta(\omega)} > 0.5$ and $V_{\rm noise} < 1$ simultaneously in both cases.
We note that similar figure of merit has recently been developed by Coudreau {\it et al.} \cite{coudreau}, during the course of this work.

\begin{figure}
\begin{center}
\includegraphics[width=\columnwidth]{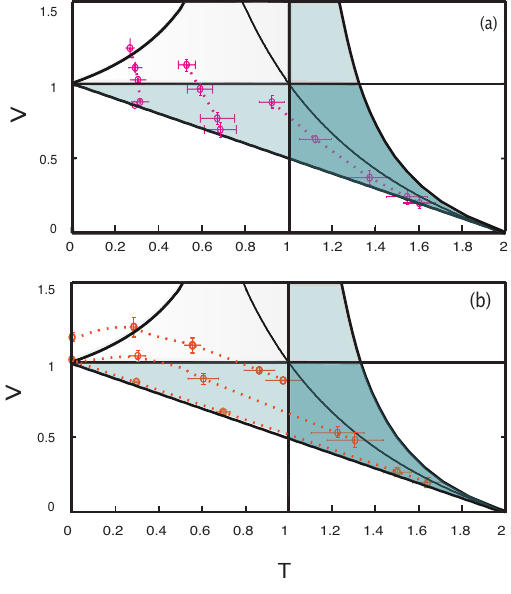}
\caption{TV diagrams showing the performance of the EIT memory. (a) shows the evolution of its efficiency for three different $\gamma_0$ values, the dotted lines representing the loci for a constant $\gamma_0$ and varying $\gamma_c$. 
 (b) shows the evolution the EIT memory for three different $\gamma_c$ values, the dotted lines representing the loci for a constant $\gamma_c$ and varying $\gamma_0$.}
\label{TVnumerics}
\end{center}
\end{figure}

Using the numerical model presented in the earlier sections, we investigate the parameters required to implement the storage of optical information in the quantum regime.  We model the situation where a fixed medium of length $L=12\,$cm with an atomic density of $10^{12}\,$cm$^{3}$ is used to store an optical signal encoded on a pulse. The length of the pulse is chosen here to be $50/\gamma$ and the information is encoded on the quadrature amplitudes at a sideband frequency $\omega=0.005\gamma$. 
We plot the evolution of the EIT-performance as a function of the decoherence rates in Fig.~\ref{TVnumerics}.
At zero decoherence rates, we note that $(T,V)$ is not (2,0) because of the finite optical thickness of the EIT medium. 
With the above parameters the bounds in Equation (\ref{inequality}) are satisfied by only one order of magnitude which makes the storage process non ideal even in the absence of dephasing. 
The no-cloning limit is however still beaten in that case.
We see that the evolution of the performances of the memory with $\gamma_0$ and $\gamma_c$ is radically different as predicted earlier.
When the decoherence rates increase, $T$ monotonically decreases and at some decoherence rate value the quantum regime is not reached anymore in both cases.
It is also important to mention that for any value of the couple ($\gamma_0$,$\gamma_c$) the region
D is never reached. The gain term from population exchange is always associated with loss so the EIT memory can never get close to the lossless amplifier regime.

We also wish to stress the difference between our figure of merit and the one used in \cite{Gor}. 
The present figure of merit considers the signal to noise ratios for both quadratures at a given sideband frequency, as well as the amount
of excess noise added to both quadratures of the field. The ratio between the total number of photons between the output and input also provides a 
figure of merit for a quantum memory that does not introduce uncorrelated extra photons in the light field output mode. 



\section{Conclusion}
We have developed a quantum multi-mode treatment describing the storage of the quantum information encoded on the sideband quadrature amplitudes of a light pulse using both stochastic simulations and an analytical treatment for EIT.  The two models included both the atomic noise and decoherence rates of realistic experiments.  In our model we have, however, assumed an ideal three level atomic structure with incident light fields that have constant transverse spatial intensities, and a mono-kinetic
atomic ensemble for which light is exactly tuned on resonance.  
We have also neglected the effect of the back coupling from spontaneous emission into the light field, such as ``radiation trapping" \cite{matsko3}.  
With these assumptions, the optimum sideband frequency for which the storage process can be efficiently performed depends mainly on the optical density and the coupling beam power chosen to set-up the EIT.  
We have also calculated the time-bandwidth product of the EIT memory and have shown that it only depends on the atomic density. 
The dependance of the decoherence rates with density was not taken into account in this paper, but would therefore constitute an important figure of merit for quantum memories.

We finally proposed the use of quantum information criteria to benchmark the performance of quantum memories against an optimal classical measure-and-prepare scheme. We have shown that for typical decoherence rates in current experiments quantum information on the sideband quadrature amplitudes can be stored for milliseconds in the no-cloning regime, in the presence of small amounts of linear loss and excess noise. 

\
\section{Acknowledgment}

We thank O. Gl\"{o}ckl, M. T. L. Hsu, B. C. Buchler, J. J. Longdell, and H. -A. Bachor for useful discussions.  
We acknowledge financial support from the Australian Research Council Centre of Excellence scheme.

\section{APPENDIX A}

We list here the stochastic equations describing the evolution of the atomic c-numbers in the presence of a quantized probe field and a classical pump for one slice $\delta z$ of the medium. 
The c-numbers $\alpha$ and $\beta$ represent the operators $\hat{\mathcal{E}}^{\dagger}(z,t)$ and $\hat{\mathcal{E}}(z,t)$.
The atomic variables $(\sigma_{3},\sigma_{4},\sigma_{5},\sigma_{6},\sigma_{7},\sigma_{9},\sigma_{10},\sigma_{11})$ represent the operators 
$(\hat{\sigma}_{13},\hat{\sigma}_{23},\hat{\sigma}_{12},\hat{\sigma}_{11},\hat{\sigma}_{22},\hat{\sigma}_{21},\hat{\sigma}_{32},\hat{\sigma}_{31})$.
The equation for $\sigma_{8}$ linearly depends on $\sigma_{6}$ and $\sigma_{7}$ via the population preservation equation $\sigma_{6}+\sigma_{7}+\sigma_{8}=1$ and is
therefore not computed. The noise terms $n_{j}$ (for $j=1$ to 18) are all delta-correlated and follow a Gaussian distribution, and have been normalized by $\frac{1}{\sqrt{n\mathcal{A}}}$. The variables $\overline{g},\overline{\gamma_{0}}$, $\overline{\gamma_{c}}$ and $\overline{E_{c}}$ are all normalized to the spontaneous emission rate $\gamma$.

\begin{widetext}
\begin{eqnarray}
	  \dot{\sigma}_{3} &=& -(1 + \overline{\gamma}_{0}/2+\overline{\gamma}_{c}/2)\sigma_{3} + \overline{E}_{c}\sigma_{5} - \alpha(1 - 2\sigma_{6} - \sigma_{7})+ \sqrt{\frac{\overline{g}}{2 \gamma}}(\alpha/\overline{g} - \sigma_{3})n_1 + i\sqrt{\frac{\overline{g}}{2 \gamma}}(\alpha/\overline{g} + \sigma_{3})n_2  \nonumber \\
	   & &- (\alpha\sigma_{4} + \overline{E}_{c}\sigma_{3})(n_3 - in_4) + \frac{1}{2\sqrt{\gamma}}\sqrt{(\overline{\gamma}_{c}+\overline{\gamma}_{0}/2)(1 - \sigma_{7} - \sigma_{6})}(i n_7 + \sqrt{2}n_{10} + i n_{12})\nonumber \\
	   & & + \sqrt{2}(\overline{\gamma}_{c}+\overline{\gamma}_{0}/2)\sigma_{4}(n_{14} - i n_{13})\nonumber \\
	 \dot{\sigma}_{4} &=& -(1 + \overline{\gamma}_{0}/2+\overline{\gamma}_{c}/2)\sigma_{4} + \alpha\sigma_{9} + \overline{E}_{c}(\sigma_{6} + 2 \sigma_{7} - 1) + \sqrt{\frac{\overline{g}}{2 \gamma}}(\sigma_{4} - \overline{E}_{c}/\overline{g}) n_1 - i\sqrt{\frac{\overline{g}}{2 \gamma}}(\sigma_{4} + \overline{E}_{c}/\overline{g}) n_2  \nonumber \\ 
	 & &+ (n_3 + i n_4)/\gamma + (\overline{\gamma}_{c}+\overline{\gamma}_{0}/2)(1 - \sigma_{6} - \sigma_{7}) (in_{15} + n_{16})/4 \nonumber \\
	  \dot{\sigma}_{5} &=& -(\overline{\gamma}_{c}+\overline{\gamma_{0}})\sigma_{5} - \alpha\sigma_{10} - \overline{E}_{c}\sigma_{3} - \sqrt{\frac{\overline{g}}{2 \gamma}}\sigma_{5}(n_1 - in_2) + ( \alpha(\sigma_{6} - \sigma_{7}) + (\overline{\gamma}_{c}+\overline{\gamma}_{0}/2)\sigma_{3} )(n_3 - in_4)/2 \nonumber \\
	  & &+ \frac{1}{2\sqrt{2}\gamma}(n_5 - in_6)+ \frac{1}{2\sqrt{\gamma}}\sqrt{\alpha\sigma_{11} + \beta\sigma_{3} + 1 - \sigma_{6} - \sigma_{7} +2\overline{\gamma}_{0}\sigma_{7}+ \overline{\gamma}_{c}(\sigma_{6} + \sigma_{7}})(i n_8 + \sqrt{2}n_9 + in_{11}) \nonumber \\
		   \dot{\sigma}_{6} &=& 1 - \sigma_{6} - \sigma_{7} - \overline{\gamma}_{c}(\sigma_{6} - \sigma_{7}) - \alpha\sigma_{11} - \beta\sigma_{3} - \frac{\overline{\gamma}_{c}}{\sqrt{2\overline{g}\gamma}}(n_1 + i n_2) -\alpha\sigma_{9}(n_3 - in_4)/2 \nonumber \\
		   & &+ \sqrt{\alpha\sigma_{11} + \beta\sigma_{3} + 1 - \sigma_{6} - \sigma_{7}}\sqrt{\alpha\sigma_{11} + \beta\sigma_{3} + 1 - \sigma_{6} - \sigma_{7} + \overline{\gamma}_{c}(\sigma_{6} + \sigma_{7})} (n_5 + in_6)\nonumber \\
		   & & + \sqrt{\frac{\overline{\gamma_c}}{2\gamma}(\sigma_{6} + \sigma_{7}})(n_7 -n_{12})- \frac{1}{\sqrt{\gamma}}\sqrt{\alpha\sigma_{11} + \beta\sigma_{3} + 1 - \sigma_{6} - \sigma_{7}}n_9\nonumber \\
		   & & + \sqrt{\alpha\sigma_{11} + \beta\sigma_{3} + 1 - \sigma_{6} - \sigma_{7}}\sqrt{\alpha\sigma_{11} + \beta\sigma_{3} + 1 - \sigma_{6} - \sigma_{7} + \overline{\gamma}_{c}(\sigma_{6} + \sigma_{7})+\overline{\gamma}_{0}\sigma_{7}}(n_{14} - in_{13})\nonumber \\
		   & & - \beta\sigma_{5}(in_{15} + n_{16})/2 - \frac{\overline{\gamma}_{c}}{\gamma}(n_{18} - in_{17})\nonumber \\   
  	  \dot{\sigma}_{7}&=& 1 - \sigma_{6} - \sigma_{7} - \overline{\gamma}_{c}(\sigma_{7} - \sigma_{6}) - \overline{E}_{c}(\sigma_{4} + \sigma_{10}) + \frac{\overline{\gamma}_{c}}{\sqrt{2\overline{g} \gamma}}(n_1 + in_2) + \alpha\sigma_{9}(n_3 - in_4)/2 \nonumber \\
	  & &+ \sqrt{2}(\alpha\sigma_{10} + \overline{E}_{c}\sigma_{3})(n_5 + in_6) + \sqrt{\frac{\overline{\gamma}_{c}}{2\gamma}(\sigma_{6} + \sigma_{7})}(n_{12} - n_7) + \frac{1}{\sqrt{2\gamma}}\sqrt{\overline{E}_{c}(\sigma_{4} + \sigma_{10}) + 1 - \sigma_{6} - \sigma_{7}}(n_8 - n_{11}) \nonumber \\
	  & &+ \sqrt{2}(\beta\sigma_{4} + \overline{E}_{c}\sigma_{11})(n_{14} - i n_{13}) + \beta\sigma_{5}(n_{16} + in_{15})/2 + \frac{\overline{\gamma}_{c}}{\sqrt{2\overline{g}\gamma}}(n_{18} - i n_{17}) \nonumber \\
	  \dot{\sigma}_{9} &=& -(\overline{\gamma}_{c}+\overline{\gamma}_{0})\sigma_{9} - \beta\sigma_{4} - \overline{E}_{c}\sigma_{11} + \frac{1}{2\sqrt{\gamma}}\sqrt{\alpha\sigma_{11} + \beta\sigma_{3} + 1 - \sigma_{6} - \sigma_{7} + \overline{\gamma}_{c}(\sigma_{6} + \sigma_{7})+\overline{\gamma}_{0}\sigma_{7}}(-in_{8} + \sqrt{2}n_9 - in_{11})\nonumber \\
	  & & + \frac{1}{2\sqrt{2}\gamma}(n_{14} + i n_{13}) + (\beta(\sigma_{6}-\sigma_{7}) + (\overline{\gamma}_{c}+\overline{\gamma}_{0}/2)\sigma_{11})(i n_{15} + n_{16})/2 - \sqrt{\frac{\overline{g}}{2\gamma}}\sigma_{9}(i n_{17} + n_{18})\nonumber \\
	  \dot{\sigma}_{10} &=& -(1 + \overline{\gamma}_{0}/2+\overline{\gamma}_{c}/2)\sigma_{10} + \beta\sigma_{5} + \overline{E}_{c}(2\sigma_{7} + \sigma_{6} - 1) +(\overline{\gamma}_{c}+\overline{\gamma}_{0}/2)(1- \sigma_{6} - \sigma_{7})(n_3 - in_4)/4 + (n_{16} - in_{15})/\gamma \nonumber \\
	  & &+ i \sqrt{\frac{\overline{g}}{2\gamma}}(\sigma_{10} + \overline{E}_{c}/\overline{g})n_{17} +  \sqrt{\frac{\overline{g}}{2\gamma}}(\sigma_{10} - \overline{E}_{c}/\overline{g})n_{18}\nonumber \\
	  \dot{\sigma}_{11} &=& -(1 + \overline{\gamma}_{0}/2+\overline{\gamma}_{c}/2)\sigma_{11} + \overline{E}_{c}\sigma_{9} - \beta(1 - 2\sigma_{6} - \sigma_{7}) + \sqrt{2}(\overline{\gamma}_{0}/2+\overline{\gamma}_{c})\sigma_{10}(n_5 + in_6)\nonumber \\
	  & &+ \frac{1}{2\sqrt{\gamma}}\sqrt{(\overline{\gamma}_{c}+\overline{\gamma}_{0}/2)(1 - \sigma_{6} - \sigma_{7})}(-i n_7 + \sqrt{2}n_{10} -i n_{12}) \nonumber \\
	  & &- (\beta\sigma_{10} + \overline{E}_{c}\sigma_{11})(i n_{15} + n_{16}) -i\sqrt{\frac{\overline{g}}{2\gamma}}(\beta/\overline{g} + \sigma_{11})n_{17} + \sqrt{\frac{\overline{g}}{2\gamma}}(\beta/\overline{g} - \sigma_{11})n_{18}.
\end{eqnarray}
\end{widetext}

\section{APPENDIX B}

In this Appendix, we list the non-zero Langevin correlations corresponding to the system of equations (\ref{max}).

\begin{eqnarray}
\langle \tilde{F}_{13}(z_1, \omega_1)
\tilde{F}_{13}^{\dagger}(z_2, \omega_2) \rangle & = &
\frac{\delta(z_1 - z_2) \delta(\omega_1 +
\omega_2)}{n \mathcal{A}} \nonumber \\
\times ( (\gamma + \gamma_c+\gamma_0)\langle \hat{\sigma}_{33} \rangle & + & 2
\gamma \langle \hat{\sigma}_{11} \rangle - \gamma_{c}
\langle \hat{\sigma}_{11} - \hat{\sigma}_{22} \rangle )
\nonumber \\
\langle \tilde{F}_{13}^{\dagger}(z_1, \omega_1)
\tilde{F}_{13}(z_2, \omega_2) \rangle & = &
\frac{\delta(z_1 - z_2) \delta(\omega_1 +
\omega_2)}{n \mathcal{A}} \nonumber \\
\times ( 2\gamma\langle \hat{\sigma}_{33} \rangle & - & 2
(\gamma +\gamma_0 +\gamma_c)\langle \hat{\sigma}_{33} \rangle )
\nonumber \\
\langle \tilde{F}_{13}^{\dagger}(z_1,\omega_1)
\tilde{F}_{12}(z_2,\omega_2) \rangle & = & \frac{\delta(z_1 -
z_2) \delta(\omega_1 + \omega_2)}{n \mathcal{A} }  \nonumber \\
& \times & (\gamma_{c}+\gamma_0) \langle \hat{\sigma}_{32} \rangle
\nonumber \\
\langle \tilde{F}_{12}^{\dagger}(z_1,\omega_1)
\tilde{F}_{13}(z_2,\omega_2) \rangle & = & \frac{\delta(z_1-z_2)
\delta(\omega_1 + \omega_2)}{n \mathcal{A} }  \nonumber \\
& \times &  (\gamma_{c}+\gamma_0) \langle \hat{\sigma}_{23} \rangle \nonumber \\
\langle \tilde{F}_{12}(z_1,\omega_1)
\tilde{F}_{12}^{\dagger}(z_2,\omega_2) \rangle & = &
\frac{\delta(z_1-z_2) \delta(\omega_1 + \omega_2)}{n \mathcal{A}}\nonumber \\
 \times ((\gamma+\gamma_c+\gamma_0) \langle \hat{\sigma}_{33} \rangle & + &
\gamma_{c} \langle \hat{\sigma}_{22} + \hat{\sigma}_{11} \rangle+2\gamma_0 \langle\sigma_{11}\rangle
) \nonumber \\
\langle \tilde{F}^{\dagger}_{12}(z_1,\omega_1)
\tilde{F}_{12}(z_2,\omega_2) \rangle & = & \frac{\delta(z_1-z_2) 
\delta(\omega_1 + \omega_2)}{n \mathcal{A}}  \nonumber \\
 \times ( \gamma \langle \hat{\sigma}_{33} \rangle  + 
\gamma_{c} \langle \hat{\sigma}_{22} &+& \hat{\sigma}_{11} \rangle +2\gamma_0 \langle\sigma_{22}\rangle
). 
\end{eqnarray}

\end{document}